\documentclass{aa}  

\usepackage{caption}
\usepackage{graphicx}
\usepackage{txfonts}
\usepackage{longtable}
\usepackage{threeparttablex}
\usepackage{natbib}
\usepackage{float}
\usepackage{enumitem}
\usepackage{subcaption}
\usepackage{mwe}
\usepackage{bm}
\usepackage{multirow}
\usepackage[switch]{lineno}
\usepackage{hyperref}
\hypersetup{
  colorlinks=true,   
  allcolors=blue
}
\usepackage{orcidlink}

\makeatletter
\newif\if@cont
\def\caption{\refstepcounter\@captype \@dblarg{\@caption\@captype}}
\def\contcaption{\@conttrue\SFB@caption\@captype}

\long\def\SFB@caption#1#2{%
 \begingroup
  \@parboxrestore
  \reset@font\normalsize
  \@makecaption{\csname fnum@#1\endcsname}{\ignorespaces #2}\par
 \endgroup}
\makeatother

\newcommand{\mspplim}{60\,ms}

\newcommand{\maspy}{$\mathrm{mas~yr^{-1}}$}

\newcommand{\kmps}{$\mathrm{km~s^{-1}}$}
\newcommand{\multilinecomment}[1]{}

\usepackage{color}

\begin{document}

   \title{The systemic recoil velocity distribution and the scale height of field millisecond pulsar systems: Implications on neutron star retention fractions in star clusters}

   \titlerunning{The systemic recoil velocities and the scale height of field millisecond pulsar systems}

   \author{Hao Ding
          \inst{1}\fnmsep\thanks{EACOA Fellow}\orcidlink{0000-0002-9174-638X}
          }

   \institute{Mizusawa VLBI Observatory, National Astronomical Observatory of Japan, 2-12 Hoshigaoka-cho, Mizusawa, Oshu, Iwate 023-0861, Japan\\
              \email{hdingastro@hotmail.com}
             }

   \date{}

 
  \abstract
   {}
   {The systemic recoil velocity ($v_\mathrm{sys}$) distribution of millisecond pulsars (MSPs) is essential for understanding the MSP formation channel(s) and for estimating the retention fractions of MSPs in star clusters, which can potentially be determined with precise astrometry of MSPs. However, the determination is complicated by MSPs' long-term dynamic evolution and the scarcity of radial velocity measurements. This work looks to overcoming the complexity and deriving the MSP $v_\mathrm{sys}$ distribution from high-precision astrometric measurements.}
   {We compiled 64 field MSP systems (including 52 binary MSPs and 12 solitary MSPs) that are well astrometrically determined, and calculated their transverse peculiar (or space) velocities $\boldsymbol{v}_\perp$ and Galactic heights $z$. Assuming that the Galactic-longitude components $v_\mathrm{l}$ of $\boldsymbol{v}_\perp$ are statistically stable over time (the ``stable-$v_\mathrm{l}$'' assumption), we approached the distribution of the $v_\mathrm{l}$ components of $\boldsymbol{v}_\mathrm{sys}$ by the observed $v_\mathrm{l}$ sample. 
   Under the ``isotropic-$\boldsymbol{v}_\mathrm{sys}$'' assumption that $\boldsymbol{v}_\mathrm{sys}$ directions are uniformly distributed, we derived the MSP $v_\mathrm{sys}$ distribution from the distribution of the $v_\mathrm{l}$ component of $\mathrm{v}_\mathrm{sys}$. Based on the derived $v_\mathrm{sys}$ distribution, we tested the ``stable-$v_\mathrm{l}$'' assumption with dynamical population synthesis (DPS). Additionally, by matching the observed $z$ and the Galactic-latitude components $v_\mathrm{b}$ of $\boldsymbol{v}_\perp$ to the DPS counterparts, we estimated the initial and the current Galaxy-wide scale heights of field MSP systems.} 
   {We find that solitary field MSPs have similar $v_\mathrm{l}$ magnitude as binary ones. Additionally, the observed $v_\mathrm{l}$ can be well described by a linear combination of three normal distributions. Accordingly, the MSP $v_\mathrm{sys}$ distribution can be approximated by a linear combination of three Maxwellian components. 
   Our DPS analysis verified the ``stable-$v_\mathrm{l}$'' assumption in the parameter space of this work, and estimated the initial and the current Galaxy-wide scale heights of field MSP systems to be about 0.32\,kpc and 0.68\,kpc, respectively.  
   }
   {According to the MSP $v_\mathrm{sys}$ distribution, $\approx14$\% of all the MSPs born in a globular cluster with the nominal 50\,\kmps\ central escape velocity can be retained.  
   Therefore, the $v_\mathrm{sys}$ distribution of field MSP systems may account for the high number of MSPs discovered in globular clusters, which implies that MSPs in star clusters may follow the same formation channel(s) as field MSP systems.}


   \maketitle
%

\section{Introduction}
\label{sec:intro}

Compact stars are normally the dense remnants of stars that have exhausted their nuclear fuel, resulting in objects such as white dwarfs (WDs), neutron stars (NSs), and black holes (BHs). Distinct formation mechanism(s) have been proposed for different types of compact stars. For example, stellar BHs are believed to form mainly through either direct collapse (of massive stars) or a delayed collapse \citep[e.g.][]{Mirabel17}. In the latter scenario, a progenitor star's core, following a supernova explosion, collapses into a proto-NS first, then succumbs to further collapse into a BH due to fallback accretion \citep[e.g.][]{Zhang08}.
As another major form of compact stars, NSs can be born with core-collapse supernovae (CCSNe), while having alternative formation channels such as accretion-induced collapse (AIC; e.g. \citealp{Tauris13}) and WD-WD mergers \citep[e.g.][]{Levan06}. 
Moreover, NSs encompass various subgroups such as magnetars and millisecond pulsar (MSP) systems, with each NS subgroup having its primary and alternative formation theories \citep[e.g.][]{Ding22}.

As distinct formation channels of compact stars may lead to different natal kicks imparted to the compact star, kinematic investigation of compact stars holds a key to discriminating between different formation channels. This investigation can be realized with precise astrometry using space-based telescopes operating at optical/infrared or through very long baseline interferometry (VLBI) observations made at radio, which has been pursued in the study of stellar BHs \citep[e.g.][]{Mirabel02,Willems05,Atri19} and various NS subgroups \citep[e.g.][]{Tendulkar13,Gonzalez11,Ding23,Ding24,Ding24b,Disberg24}.

In practice, the birth velocity and position of a compact star or a compact star system (i.e., a gravitationally bound system of $\lesssim3$ stars hosting a compact star) can be derived with the knowledge of {\bf 1)} the observed 3-dimensional (3D) position, {\bf 2)} the observed 3D velocity and {\bf 3)} the age of the system, by integrating the motion backward over time assuming a model of Galactic potential. 
However, the age of a compact star system is usually not well constrained, therefore requiring additional assumption about the birth site (e.g. thin Galactic disk) of the compact star system \citep[e.g.][]{Fragos09,Atri19}.
Furthermore, most compact star systems, especially pulsar systems, do not have the information of line-of-sight radial velocities $v_\mathrm{r}$. 
As a result,   
the investigation of the systemic recoil velocities $\boldsymbol{v}_\mathrm{sys}$ (the recoil velocities of lone compact stars or stellar systems hosting compact stars) of pulsar systems has been mainly limited to kinematically young ($\lesssim10$\,Myr old) systems \citep[e.g.][]{Cordes98,Kramer03,Lorimer06,Verbunt17,Igoshev20,Ding24b} in order to minimize the dynamic evolution (of the compact star systems) through the Galactic potential.
We note that $v_\mathrm{r}$ are distinct from the so-called ``Galactic radial velocity'' $v_\mathrm{R}$ defined as the radial component of motion in cylindrical Galactocentric coordinates. Unlike $v_\mathrm{r}$, $v_\mathrm{R}$ can be precisely inferred from the transverse velocity of a compact star when the line of sight (to the compact star) is nearly perpendicular to the star's radial vector in cylindrical Galactocentric coordinates.

On the other hand, for any specific subgroup of kinematically old ($\gtrsim40$\,Myr) compact star systems that do not have $v_\mathrm{r}$ information, the observed transverse peculiar velocities $\textit{\textbf{v}}_\perp$ (also referred to as transverse space velocity, i.e., velocity component tangential to the line of sight, with respect to the standard of rest of the neighborhood) is expected to deviate from the $v_\mathrm{sys}$ component tangential to the sightline \citep[e.g.][]{Disberg24a}.
Despite the deviation, it has been proposed by \citet{Faucher-Giguere06} that the Galactic-longitude component $v_\mathrm{l}$ of $\boldsymbol{v}_\perp$ is statistically stable over time, which serves as a critical assumption that can link $v_\mathrm{l}$ to the $v_\mathrm{sys}$ distribution. For this assumption to be validated in the parameter space of interest, dynamical population synthesis (DPS) is required, which simulates the motion of stellar objects through the Galactic potential assuming initial states (i.e., positions and velocities) \citep[e.g.][]{Cordes97}.

Promising supreme timing stability, MSPs have been extensively studied, and are used to test gravitational theories \citep[e.g.][]{Freire12,Zhu19,Freire24} and detect gravitational-wave background at the nano-Hz regime \citep[e.g.][]{Agazie23a,Antoniadis23a,Reardon23,Xu23,Miles25}.
With spin periods of only $\lesssim~$\mspplim, MSPs are widely believed to have been spun up during the accretion from their respective donor stars. However, as explained in Section~1.3.1 of \citet{Ding22}, the unexpectedly large number of MSPs retained in globular clusters \citep[e.g.][]{Pfahl02} and possibly also at the Galactic centre \citep[e.g.][]{Abazajian12} suggest very small $v_\mathrm{sys}$ of MSPs \citep[e.g.][]{Boodram22}, which reinforces exotic MSP formation channels such as the AIC channel \citep[e.g.][]{Gautam22}. 
Accordingly, it is likely that the MSP $v_\mathrm{sys}$ distribution is multi-modal and/or multi-component (see Section~\ref{sec:v_sys_distribution} or Appendix~\ref{subap:candidate_vl_PDFs} for explanation), which has not been indicated by previous studies of MSP kinematics likely due to the relatively large astrometric uncertainties \citep{Hobbs05,Gonzalez11}.

Up till now, a well defined MSP $v_\mathrm{sys}$ distribution remains largely absent, mainly due to the aforementioned technical challenges (i.e., long-term dynamic evolution and the lack of $v_\mathrm{r}$ information) and relatively large astrometric uncertainties. This absence hinders the studies of MSP formation channels, and results in weak constraints on the NS retention fractions in star clusters.
Recent years have seen a major increase in the number of well astrometrically determined MSPs \citep[e.g.][]{Ding23,Shamohammadi24}, which promises a refined characterization of the MSP $v_\mathrm{sys}$ distribution. 
In this paper, we determine the MSP $v_\mathrm{sys}$ distribution with the observed $v_\mathrm{l}$ calculated for 64 well astrometrically determined MSPs, then discuss its implications for the NS retention fractions in star clusters. 
Throughout this paper, uncertainties are stated at 68\% confidence.

\section{The observed transverse peculiar velocities and Galactic heights of millisecond pulsars}
\label{sec:obs}

The establishment of the observed $\boldsymbol{v}_\perp$ distribution requires precise determination of both proper motions and distances for the MSPs, especially as MSP $v_\perp$ are much lower than those of normal pulsars \citep{Hobbs05,Gonzalez11,Matthews16,Shamohammadi24}. 
Applying low-precision astrometric results to MSP kinematics studies would not only shift the average $v_\perp$ to the higher end, but also reduce the ability to resolve the potential multiple modes in the $\boldsymbol{v}_\perp$ distribution.
Compared to other methods, the determination of trigonometric parallaxes through VLBI astrometry \citep[e.g.][]{Ding23}, pulsar timing \citep[e.g.][]{Perera19,Shamohammadi24}, or Gaia astrometry \citep[e.g.][]{Jennings18,Moran23}, offers the most precise and reliable distances for MSPs.
Recent releases of the astrometric results by \citet{Ding23} and \citet{Shamohammadi24} have significantly enriched the sample of well astrometrically determined MSPs, which offers a rare opportunity to make statistically significant investigation of the MSP kinematics based purely on well astrometrically determined MSPs.

We searched for viable MSPs within the {\tt PSRCAT} Catalogue\footnote{\label{footnote:psrcat}\url{https://www.atnf.csiro.au/research/pulsar/psrcat/}} \citep{Manchester05}. 
Due to relatively frequent dynamic interactions inside globular clusters, 
the $\boldsymbol{v}_\perp$ of globular cluster MSPs are likely not reflective of their $v_\mathrm{sys}$. With globular cluster MSPs excluded from the study, we reached 64 field MSP systems having significant ($>3\,\sigma$) proper motion and parallax determinations (see Table~\ref{tab:MSPs_d__v_t}). 
When only focusing on field MSPs with $>5\,\sigma$ and $>7\,\sigma$ parallaxes, we obtained two high-precision samples of 42 and 30 MSPs, respectively (see Appendix~\ref{subap:obtain_high_precision_samples}).
The details of the sample selection are described in Appendix~\ref{ap:sample_selection}.

Following the methods detailed in Section~6 of \citet{Ding23}, we derived the distances $d$ and the $\boldsymbol{v}_\perp$ of the 64 MSPs from their parallaxes and proper motions.
The $\boldsymbol{v}_\perp$ have been corrected for the Solar motion and converted to the Local Standard of Rest of each pulsar.
In deriving $d$ from the parallaxes, the Lutz-Kelker bias \citep{Lutz73} has been systematically corrected. This correction is particularly important for MSPs with relatively low parallax significances ($\lesssim6\,\sigma$), as directly inverting parallaxes tends to underestimate their $d$ (and likely $v_\perp$ as well).
From the $d$, we calculated the Galactic heights $z\equiv d\sin{b}$ for the 64 MSPs, where $b$ refer to Galactic latitudes. The derived $d$, $\boldsymbol{v}_\perp$ and $z$ are summarized in Table~\ref{tab:MSPs_d__v_t}, where the columns related to $\boldsymbol{v}_\perp$ include the $\boldsymbol{v}_\perp$ magnitudes $v_\perp$, the Galactic-longitude component $v_\mathrm{l}$ and the Galactic-latitude component $v_\mathrm{b}$ of $\boldsymbol{v}_\perp$. 
On top of $v_\mathrm{l}$ and $v_\mathrm{b}$, $z$ serves as an additional indicator of MSP kinematics, as the average level of $|z|$ expected to grow with higher $v_\mathrm{sys}$.
Therefore, this work includes the studies of three samples of measurements: the $v_\mathrm{l}$, the $v_\mathrm{b}$, and the $z$ samples. 

Based on the calculations of the Pearson coefficients, no linear correlation is found between any two of the three samples. Accordingly, the three samples are treated independently.
We conducted Wilcoxon signed-rank test on each of the three samples of 64 measurements, and confirmed no indication of asymmetry about 0 in any sample. Accordingly, in the following analysis, we assumed that the samples of the observed $v_\mathrm{l}$, $v_\mathrm{b}$ and $z$ are symmetric about 0. 

\begin{figure*}[h]
    \centering    \includegraphics[width=0.98\textwidth]{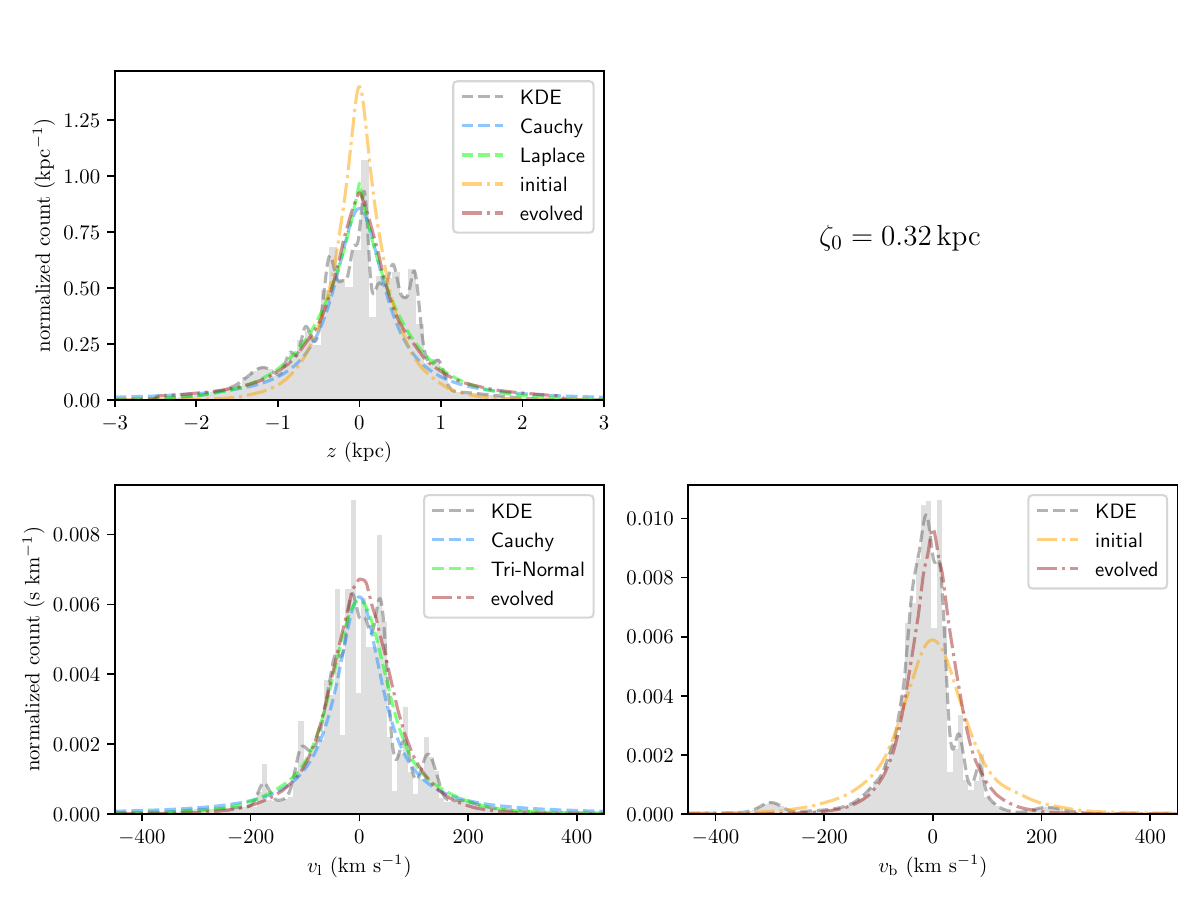}
    \caption{
    The upper-left, lower-left and lower-right panels display the histograms of Galactic heights $z$, the Galactic-longitude component $v_\mathrm{l}$ and the Galactic-latitude component $v_\mathrm{b}$ of transverse peculiar velocities, respectively. Each histogram is concatenated from 10000 simulations drawn from the assumed split normal distributions for the 64 measurements (of $z$, $v_\mathrm{l}$ or $v_\mathrm{b}$; see Table~\ref{tab:MSPs_d__v_t}). The gray dashed curves show the normalized count smoothed with kernel density estimation (using {\tt scipy.stats.Gaussian\_kde}). The best-fit Cauchy, Laplace and three-component normal distributions are plotted in blue and green dashed curves. Overlaid are the dynamical population synthesis results (labeled as ``evolved'') that best match the observed $z$, $v_\mathrm{l}$ and $v_\mathrm{b}$, based on the determined initial MSP scale height $\zeta_0=0.32$\,kpc. The corresponding initial distributions are also provided except for $v_\mathrm{l}$, as the initial $v_\mathrm{l}$ distribution adopted in the dynamical population synthesis is identical to the best-fit three-component normal distribution (a linear combination of three normal distributions).}
    \label{fig:v_and_z_distributions}
\end{figure*}

To {\bf 1)} visualize the three samples of measurements, and {\bf 2)} investigate the underlying probability density function (PDF) for each sample, we assumed each measurement follows a split normal distribution, simulated 10,000 values per measurement, and concatenated the simulations into a chain of 640,000 values for each of $v_\mathrm{l}$, $v_\mathrm{b}$ and $z$. 
Despite the simulated sample of 640,000 following essentially a multi-modal distribution, the visualization is expected to characterize where the measurements overlap. The resultant histograms are displayed in Figure~\ref{fig:v_and_z_distributions}.
According to Table~\ref{tab:MSPs_d__v_t} and Figure~\ref{fig:v_and_z_distributions}, most of the 64 MSPs are within the scale height of the Galactic thick disk (i.e., $\sim1$\,kpc).

\subsection{Binary vs. Solitary}
\label{subsec:binary_vs_solitary}

While most field MSPs are believed to be formed in binary systems where the NSs are spun up during the accretion from their main-sequence companions \citep[e.g.][]{Alpar82,Bhattacharya91}, $\gtrsim20\%$ of field MSPs are found to have no companions. 
It is proposed that at least some of these solitary (or single) MSPs have ablated their companions after being spun up through accretion \citep[e.g.][]{Fruchter88,Kluzniak88}. 
This scenario is supported by the discoveries of planets around the MSP PSR~J1300$+$1240 (or PSR~B1257$+$12) \citep{Wolszczan92,Konacki03}, and planetary-mass companions around the MSPs PSRs~J1719$-$1438 and J2322$-$2650 \citep{Bailes11,Spiewak18}. 
Additionally, the persistent
achromatic component in the timing residuals of PSR~J1939$+$2134 (or PSR~B1937$+$21) has been attributed to the presence of an asteroid belt surrounding the MSP \citep{Shannon13}.
Of the four aforementioned MSPs supporting the ``companion ablation'' scenario, PSRs~J1300$+$1240 and J1939$+$2134 are included in this work, and are classified as solitary MSPs (see Table~\ref{tab:MSPs_d__v_t}).

In addition to the companion ablation scenario, alternative theories have been proposed for the formation of solitary field MSPs, including (but not limited to) the AIC channel \citep{Tauris13,Freire13} and the strange star scenario \citep{Jiang20a}. Particularly, the AIC channel is suggested to result in relatively small peculiar velocities \citep[e.g.][]{Gautam22}. Therefore, precise astrometry of solitary MSPs can be used to potentially distinguish between the companion ablation and the AIC channels. Previous astrometric investigations by \citet{Hobbs05,Gonzalez11,Matthews16,Shamohammadi24} suggest no indication that solitary field MSPs and binary ones follow different $v_\perp$ distributions. Here, we revisit the investigation in better detail, incorporating refined astrometric data and improved analytical techniques.

\begin{figure*}[h]
    \centering    \includegraphics[width=0.9\textwidth]{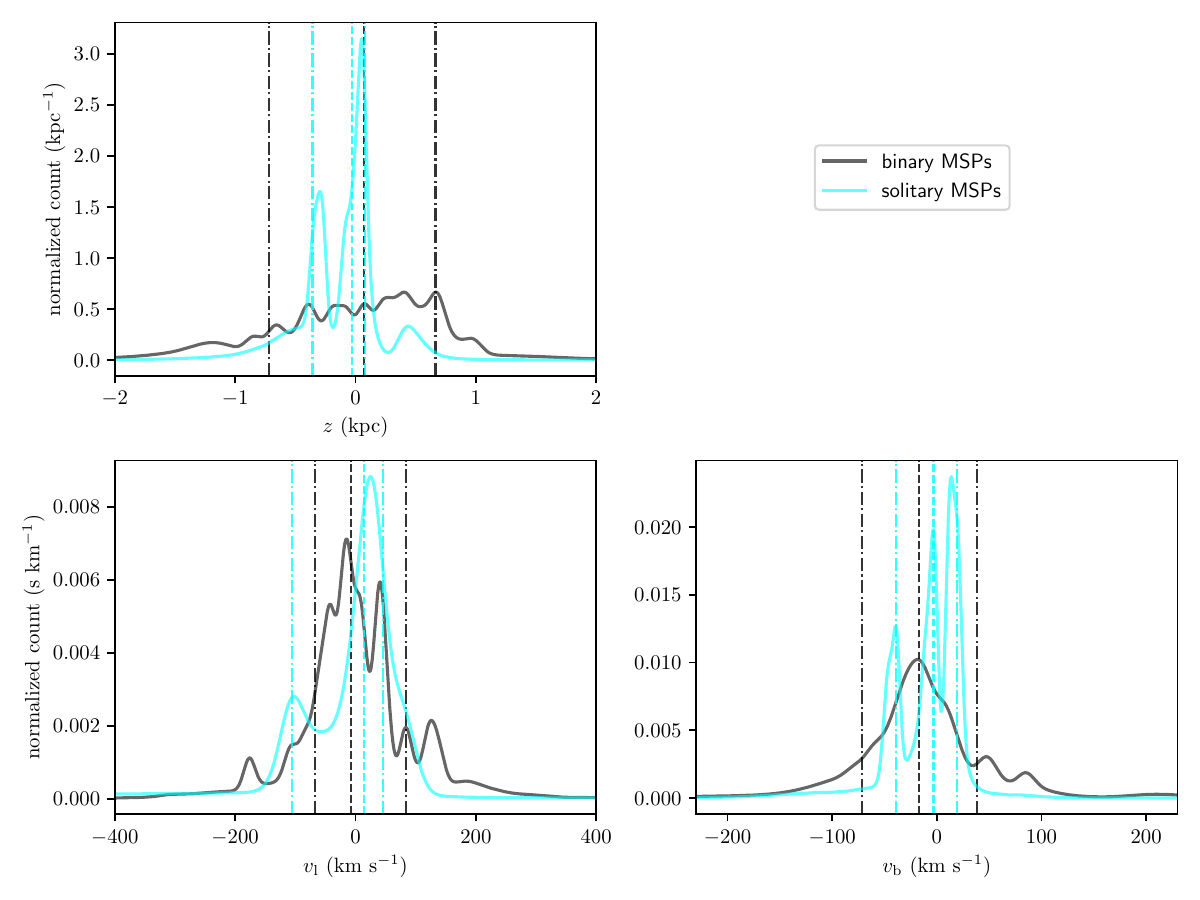}
    \caption{
    The curves in the upper-left, lower-left and lower-right panels display the normalized histograms of the observed Galactic heights $z$, Galactic-longitude component $v_\mathrm{l}$ and Galactic-latitude component $v_\mathrm{b}$ of transverse peculiar velocities, respectively, for 52 binary MSPs and 12 solitary MSPs (see Table~\ref{tab:MSPs_d__v_t}). 
    The histograms have been smoothed with kernel density estimation using {\tt scipy.stats.Gaussian\_kde}. 
    Each histogram is concatenated from 10000 simulations drawn from the assumed split normal distributions for the measurements (of $z$, $v_\mathrm{l}$ or $v_\mathrm{b}$; see Table~\ref{tab:MSPs_d__v_t}).}
    \label{fig:binary_vs_solitary}
\end{figure*}

Among the 64 selected MSPs, 12 are solitary MSPs and 52 MSPs are in binary systems (see Table~\ref{tab:MSPs_d__v_t}).
Following the aforementioned way of visualization, we display the normalized histograms of $v_\mathrm{l}$, $v_\mathrm{b}$ and $z$ for the 52 binary MSPs and 12 the solitary MSPs in Figure~\ref{fig:binary_vs_solitary}. The global magnitudes of $v_\mathrm{l}$, $v_\mathrm{b}$ and $z$ of the solitary and the binary MSPs are summarized in Table~\ref{tab:binary_vs_solitary}.
We found that the median and dispersion of $|v_\mathrm{l}|$ of solitary MSPs well agree with the counterparts of binary MSPs. Assuming {\bf 1)} the $v_\mathrm{l}$ distribution is stable over time (see Section~\ref{sec:popSyn} for justification) and {\bf 2)} MSP systemic recoil velocities are equally possible in all directions, the agreement in $|v_\mathrm{l}|$ suggests that solitary field MSPs may follow a systemic recoil velocity distribution similar to that of binary field MSPs.
Accordingly, most solitary field MSPs are expected to follow the primary formation channel of binary MSPs (in which they are spun up through accretion from their companions), except that solitary field MSPs eventually evaporate their companions through irradiation.
Though the AIC channel is suggested to produce the bulk of the slow-moving MSPs that are presumably retained in the Galactic centre and contribute to the gamma-ray excess observed in the direction of the Galactic centre \citep{Boodram22,Gautam22}, the observed $v_\mathrm{l}$ suggest that the AIC channel should contribute little to the production of solitary field MSPs. 
In the following discussions, we assume that solitary and binary field MSPs share the same systemic recoil velocity distribution, and derive the systemic recoil velocity distribution with the full sample of 64 MSPs.

On the other hand, both $|z|$ and $|v_\mathrm{b}|$ of solitary MSPs appear to be overall smaller than the counterparts of the binary MSPs. In other words, binary MSPs are found to be more spread out from the Galactic plane (than solitary MSPs), with their $|z|$ nearly following a uniform distribution between 0--1\,kpc (see Figure~\ref{fig:binary_vs_solitary}). This spread may reflect the observational bias that searching for binary MSPs in the scattered low-Galactic-latitude sky regions is disproportionately harder (compared to solitary MSPs), which could be gradually resolved with increasingly advanced binary pulsar search techniques \citep[e.g.][]{Balakrishnan22,Sengar25}. 
As the sample of solitary field MSPs remains small, we use the full sample of 64 MSPs to estimate the initial and the evolved scale heights of field MSPs in Section~\ref{sec:popSyn}, which should be considered the weighted average values of the two sub-samples (of solitary and binary MSPs).

\begin{table}[h]
     \centering
     \caption[]{\label{tab:binary_vs_solitary}
     Global magnitudes of the Galactic heights and the transverse peculiar velocities of the binary and the solitary millisecond pulsars.}
     
     {\renewcommand{\arraystretch}{1.3}
     \resizebox{0.7\columnwidth}{!}{
    \begin{tabular}{cccc}
        \hline 
        \hline

    MSPs & $|v_\mathrm{l}|$ & $|v_\mathrm{b}|$ & $|z|$ \\
     & (\kmps) & (\kmps) & (kpc) \\
     \hline
     binary & $44^{+85}_{-31}$ & $34^{+56}_{-24}$ & $0.46^{+0.51}_{-0.30}$ \\
     solitary & $40^{+69}_{-26}$ & $17^{+26}_{-13}$ & $0.13^{+0.33}_{-0.08}$ \\
     
        \hline
        
        \hline
    \end{tabular}
    }}
     \tablefoot{
    The lower bound, the estimate, and the upper bound corresponds, respectively, to the 16th, the 50th, and the 84th percentile of each of the $|v_\mathrm{l}|$, $|v_\mathrm{b}|$, and $|z|$ chains.
    }
    \end{table}

\subsection{Modal analysis based on different parallax significance}
\label{subsec:modal_analysis}

Following the method detailed in Section~7.2.3 of \citet{Ding24}, we ran Hartigan's dip tests \citep{Hartigan85} on the three samples (of 64 measurements), and found no indication of multi-modality in any of the samples. 
As the $v_\mathrm{l}$ distribution is believed to be stable over time \citep{Faucher-Giguere06}, here we mainly discuss the dip tests on the $v_\mathrm{l}$ sample. When assuming that the $v_\mathrm{l}$ sample follows a distribution symmetric about 0 (as is supported by the aforementioned Wilcoxon signed-rank tests), the $|v_\mathrm{l}|$ sample should, in principle, enhance the indicators of potential multi-modality.
Should the $v_\mathrm{l}$ sample truly follow the multi-modal distribution used in the visualization, about 180 and 240 additional $v_\mathrm{l}$ measurements would be needed to rule out uni-modality of the $|v_\mathrm{l}|$ sample at 90\% and 95\% confidence, respectively, according to our simulations. 

Additionally, we ran the same analysis on high-precision MSP samples obtained with higher thresholds of parallax significance $\varpi/\sigma_\varpi$ (see Appendix~\ref{subap:obtain_high_precision_samples}). The results for $v_\mathrm{l}$ and $|v_\mathrm{l}|$ are summarized in Table~\ref{tab:diptests}. According to Table~\ref{tab:diptests}, raising $\varpi/\sigma_\varpi$ thresholds reduces MSP sample sizes and lowers the level of unimodality of $|v_\mathrm{l}|$ or $v_\mathrm{l}$ measurements. The latter trend is likely caused by small sample effects. However, it is not unlikely that more high-precision astrometric measurements made with VLBI or pulsar timing could resolve the $v_\mathrm{l}$ distribution, and reveal multi-modality of MSP $|v_\mathrm{l}|$ with certainty in future.
In the following discussions, we only focus on the full MSP sample (of 64), and assume that the $v_\mathrm{l}$ distribution is unimodal (given no indication of multi-modality).

\begin{table}[h]
     \centering
     \caption[]{\label{tab:diptests}
     The p-values (of unimodality) of Hartigan's dip tests on MSP samples (with sample size $N_\mathrm{sample}$) obtained with various parallax significances  $\varpi/\sigma_\varpi$.}
     
     {\renewcommand{\arraystretch}{1.3}
     \resizebox{0.8\columnwidth}{!}{
    \begin{tabular}{c|c|c|c|c|c}
        \hline 
        \hline
    $\varpi/\sigma_\varpi$  & $>3$ & $>4$ & $>5$ & $>6$ & $>7$ \\
        \hline
    $N_\mathrm{sample}$ & 64 & 53 & 45 & 38 & 30 \\
    \hline
    \multicolumn{6}{c}{Hartigan's dip tests on $v_\mathrm{l}$}\\
    \hline
    p-value & 0.91 & 0.89 & 0.81 & 0.77 & 0.92 \\
    $N_{0.9}$ & $\approx490$ & $\approx360$ & $\approx220$ & $\approx150$ & $\approx200$ \\
    $N_{0.95}$ & $\approx610$ & $\approx450$ & $\approx270$ & $\approx180$ & $\approx250$ \\
    \hline
    \multicolumn{6}{c}{Hartigan's dip tests on $|v_\mathrm{l}|$}\\
    \hline
    p-value & 0.64 & 0.57 & 0.43 & 0.42 & 0.33 \\
    $N_{0.9}$ & $\approx240$ & $\approx160$ & $\approx110$ & $\approx90$ & $\approx60$ \\
    $N_{0.95}$ & $\approx300$ & $\approx200$ & $\approx130$ & $\approx110$ & $\approx70$ \\
     
        \hline
        
        \hline
    \end{tabular}
    }}
    \tablefoot{
    $N_{0.9}$ and $N_{0.95}$ refers to the total number of measurements (including the $N_\mathrm{sample}$ measurements) needed to rule out uni-modality of the $v_\mathrm{l}$ (or $|v_\mathrm{l}|$) samples at 90\% and 95\% confidence, respectively, assuming the $v_\mathrm{l}$ sample truly follow the multi-modal distribution used in the visualization (see Section~\ref{sec:obs}).
    }
    \end{table}

\section{The systemic recoil velocity distribution}
\label{sec:v_sys_distribution}

We followed \citet{Faucher-Giguere06} to approach the $v_\mathrm{sys}$ distribution with the observed $v_\mathrm{l}$ sample (see Table~\ref{sec:obs}), assuming that $v_\mathrm{l}$ stays statistically stable over the long ($\lesssim10$\,Gyr) journeys of MSPs. The validation of this ``stable-$v_\mathrm{l}$'' assumption is detailed in Section~\ref{sec:popSyn}.
We determined the $v_\mathrm{sys}$ distribution in mainly three steps.

In the first step, we identified PDF candidates based on two simplifying assumptions.
Given no indication that the observed $v_\mathrm{l}$ sample is multi-modal or asymmetric about zero (see Section~\ref{sec:obs}), we assumed that the underlying PDF of the observed $v_\mathrm{l}$ sample is both uni-modal and symmetric about zero. 
Basic PDFs that meet these two requirements include (but are not limited to) normal distributions, Laplace distributions, and Cauchy distributions. Additionally, linear combinations of these basic distributions also serve as useful candidates. In particular, a linear combination of normal distributions with different scales (hereafter referred to as a multi-component normal distribution) would conveniently correspond to a multi-component Maxwell distribution for $v_\mathrm{sys}$ (assuming that $\boldsymbol{v}_\mathrm{sys}$ is equally possible in all directions for kinematically old pulsar systems).

In the second step, we fit the parameter(s) of each candidate PDF using {\tt scipy.optimize.curve\_fit} \citep{Virtanen20}, for the PDF to approximate the curve (see Figure~\ref{fig:v_and_z_distributions}) smoothed from the normalized histogram of the measurement-based simulated $v_\mathrm{l}$ sample of 640,000 (see Section~\ref{sec:obs}), where the smoothing was conducted with kernel density estimation using {\tt scipy.stats.Gaussian\_kde} \citep{Virtanen20}.   
After discarding candidates with poor fitting results, we identified two groups of prime PDF candidates (that well match the normalized $v_\mathrm{l}$ histogram), with one group being linear combinations of Cauchy distributions and the other group being linear combinations of the normal distributions. The prime candidate PDFs are described in detail in Appendix~\ref{subap:candidate_vl_PDFs}.

\begin{figure*}[h]
    \centering    \includegraphics[width=0.65\textwidth]{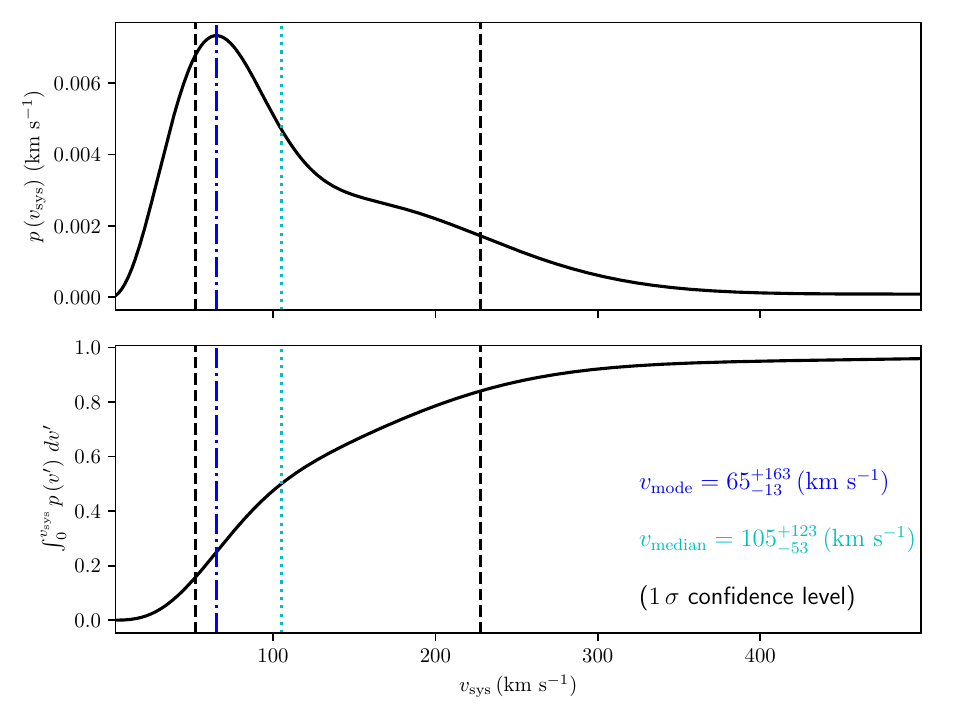}
    \caption{
    {\bf Upper:} The three-component Maxwell distribution described by Equation~\ref{eq:Vsys_distribution}, which shares the parameters of the best-fit three-component normal distribution (see Table~\ref{tab:PDF_candidates}) of the observed $v_\mathrm{l}$ sample. {\bf Lower:} The cumulative distribution function (CDF) calculated from the three-component Maxwell distribution shown in the upper panel. The dash-dotted vertical line marks the mode of the three-component Maxwell distribution, while the dashed lines and the dotted line correspond to the 16th, the 84th, and the 50th percentile of the CDF, respectively.}
    \label{fig:Vsys_PDF_CDF}
\end{figure*}

In the final step, we selected the PDF for the observed $v_\mathrm{l}$ sample by the method detailed in Appendix~\ref{subap:Bayes_analysis}. We conclude that the observed $v_\mathrm{l}$ can be well described by a Cauchy distribution with scale of 51.2\,\kmps\ (i.e., Equation~\ref{eq:vl_distribution}). Despite a three-component normal distribution being more consistent with the Cauchy distribution (see Figure~\ref{fig:v_and_z_distributions}), the more complex distribution has four more free parameters, therefore being less favorable.
On the other hand, the three-component normal distribution, described by Equation~\ref{eq:multi_scale_normal_PDF} and Table~\ref{tab:PDF_candidates}, not only serves as a good approximation to the underlying PDF of the observed $v_\mathrm{l}$ sample, but also provides a rare analytical solution for the $\boldsymbol{v}_\mathrm{sys}$ magnitude $v_\mathrm{sys}$: when assuming the $\boldsymbol{v}_\mathrm{sys}$ distribution is isotropic for kinematically old pulsar systems, $v_\mathrm{sys}$ is expected to follow a three-component Maxwell distribution (i.e., a weighted sum of three Maxwell distributions) that share the same parameters with the best-fit three-component normal distribution (see Table~\ref{tab:PDF_candidates}). 
Namely, under the ``isotropic-$\boldsymbol{v}_\mathrm{sys}$'' assumption, $v_\mathrm{sys}$ should approximately follow the PDF
\begin{equation}
\label{eq:Vsys_distribution}
\begin{split}
p=&\sqrt{\frac{2}{\pi}}\Biggl\{0.463\cdot\frac{{v_\mathrm{sys}}^2}{43.1^3}\cdot\exp{\left[-\frac{1}{2}\left(\frac{v_\mathrm{sys}}{43.1}\right)^2\right]} \\ &\quad +0.477\cdot\frac{{v_\mathrm{sys}}^2}{107.6^3}\cdot\exp{\left[-\frac{1}{2}\left(\frac{v_\mathrm{sys}}{107.6}\right)^2\right]} \\
&\quad +0.060\cdot\frac{{v_\mathrm{sys}}^2}{406.1^3}\cdot\exp{\left[-\frac{1}{2}\left(\frac{v_\mathrm{sys}}{406.1}\right)^2\right]}\Biggl\}\,\,,
\end{split}
\end{equation}
where the unit of $p$ and $v_\mathrm{sys}$ is $\mathrm{s~km^{-1}}$ and \kmps, respectively.
In contrast, no analytical solution of the $v_\mathrm{sys}$ PDF can be solved from the best-fit Cauchy distribution of $v_\mathrm{l}$. Therefore, we adopt Equation~\ref{eq:Vsys_distribution} as the $v_\mathrm{sys}$ PDF in the following analysis, which is illustrated along with its cumulative distribution function (CDF) in Figure~\ref{fig:Vsys_PDF_CDF}.

MSP velocities have been previously studied by \citet{Hobbs05,Gonzalez11,Matthews16,Shamohammadi24}, where their MSP distances were mostly estimated using observed dispersion measures (DMs) and a model \citep{Cordes02,Yao17} of the Galactic free electron distribution, as precise parallaxes were unavailable in most cases. Among these four works, \citet{Hobbs05,Gonzalez11} report the mean 3D peculiar velocity of MSPs to be $111\pm17$\,\kmps\ and $108\pm15$\,\kmps, respectively. Both values well agree with
the median $v_\mathrm{sys}$ (of field MSPs) estimated in this work (i.e., 105\,\kmps; see Figure~\ref{fig:Vsys_PDF_CDF}). 
Since no $v_\mathrm{sys}$ distribution has been derived in previous studies, a more detailed comparison with our $v_\mathrm{sys}$ distribution is not currently feasible.

Regarding the use of Equation~\ref{eq:Vsys_distribution}, we like to clarify two points.
Firstly, despite being a three-component Maxwell distribution, Equation~\ref{eq:Vsys_distribution} is still a uni-modal distribution by definition (see Figure~\ref{fig:Vsys_PDF_CDF}). Secondly, though approximating the $v_\mathrm{sys}$ PDF by Equation~\ref{eq:Vsys_distribution}, we do not suggest that field MSP systems have multiple formation channels. To further the investigation of formation channels of field MSP systems, population synthesis studies based on various MSP formation channels are needed to generate the theoretical $v_\mathrm{sys}$ PDF templates that can be compared with our observation-based counterpart.

\section{Dynamical population synthesis}
\label{sec:popSyn}

When deriving the $v_\mathrm{sys}$ PDF, we assumed $v_\mathrm{l}$ to be statistically stable over time (see Section~\ref{sec:v_sys_distribution}). We tested this ``stable-$v_\mathrm{l}$'' assumption using DPS that simulates the dynamics of field MSP systems in the Galaxy. Specifically, we like to examine if the current (or evolved) $v_\mathrm{l}$ distribution stays roughly the same as the initial $v_\mathrm{l}$ distribution.
We used Equation~\ref{eq:Vsys_distribution} as the MSP $v_\mathrm{sys}$ PDF in the DPS analysis.
The {\tt MWPotential2014} model, calculated and compiled into the {\tt galpy} package by \citet{Bovy15}, was adopted as the Galactic potential used in the DPS.
Apart from the $v_\mathrm{sys}$ PDF and the Galactic potential, two other ingredients are essential the DPS analysis --- the MSP age distribution, and the initial spatial distribution of field MSP systems in the Galaxy. The former ingredient is described in Appendix~\ref{subap:age_distribution}, while the latter is introduced as follows. 

\subsection{The initial Galactic spatial distribution}
\label{subsec:initial_spatial_distribution}

The initial Galactic distribution of field MSP systems consists of two parts --- the birth radial distribution (i.e., the spatial distribution projected to the Galactic plane) and the birth Galactic height distribution (hereafter referred to as the $z_\mathrm{i}$ distribution).
For the former part, we adopted the pulsar radial distribution established with relatively young pulsars by \citet{Xie24}, assuming that field MSPs (excluding those originating from globular clusters) are born in pulsar-forming regions similar to those of younger pulsars (such as the spiral arms and the central Galactic region; \citealp{Xie24}).

On the other hand, the $z_\mathrm{i}$ distribution of field MSP systems is largely unknown. 
As we found that the observed $z$ sample (see Table~\ref{tab:MSPs_d__v_t}) can be well described by a Laplace distribution with the scale height $\zeta^\mathrm{local}_\mathrm{obs}=0.52$\,kpc (see Figure~\ref{fig:v_and_z_distributions} and Appendix~\ref{subsubap:obs_z_PDF}), we assumed that $z_\mathrm{i}$ of Galaxy-wide field MSP systems also follow a Laplace distribution with the scale height of $\zeta_0$. 
In DPS, we realized that both the evolved $v_\mathrm{b}$ and the evolved $z$ change with $\zeta_0$ (see Appendix~\ref{subsubap:birth_scale_height}). Therefore, by matching the evolved $v_\mathrm{b}$ and the evolved $z$ to the observed counterparts, one can determine $\zeta_0$. 
The DPS procedure that determines the evolved states (including the evolved positions and the evolved velocities) is described in Section~\ref{subsec:popSyn_procedure}, while the details of the $\zeta_0$ determination are provided in Appendix~\ref{subsubap:birth_scale_height}.
We find that the DPS with $\zeta_0=0.32$\,kpc renders evolved $v_\mathrm{b}$ and evolved $z$ samples that well match the observed counterparts (see Figure~\ref{fig:z0_dependence}). Accordingly, we adopt $\zeta_0=0.32$\,kpc for the Galaxy wide $z_\mathrm{i}$ distribution of field MSP systems.

\subsection{The procedure of dynamical population synthesis}
\label{subsec:popSyn_procedure}

In DPS, we first simulated 10,000 initial positions based on the initial spatial distribution described in Section~\ref{subsec:initial_spatial_distribution}. At each position, we assumed the stellar progenitor of the MSP followed circular motion (about the axis of rotation of the Galaxy) specified by the rotation curve of the {\tt MWPotential2014} Galactic potential model. $\boldsymbol{v}_\mathrm{sys}$ were simulated based on the $v_\mathrm{sys}$ PDF (i.e., Equation~\ref{eq:Vsys_distribution}) and the ``isotropic-$\boldsymbol{v}_\mathrm{sys}$'' assumption (see Section~\ref{sec:v_sys_distribution}). By adding the simulated $\boldsymbol{v}_\mathrm{sys}$ to the position-based circular velocities, we obtained the simulated MSP velocities with respect to the Galactic centre.
Based on the simulated initial positions and velocities, we calculated the simulated initial $v_\mathrm{l}$, $v_\mathrm{b}$, and $z_\mathrm{i}$ in the same way as the calculation of the observed counterparts.
The histograms of the initial $v_\mathrm{b}$ and $z_\mathrm{i}$ samples were smoothed with kernel density estimation (using {\tt scipy.stats.Gaussian\_kde}), and overlaid to Figure~\ref{fig:v_and_z_distributions}. As the three-component normal distribution we adopt as the initial $v_\mathrm{l}$ distribution is already displayed in Figure~\ref{fig:v_and_z_distributions}, we do not overlay the smoothed histogram of the initial $v_\mathrm{l}$ sample.  

Subsequently, we performed orbit integrations with {\tt galpy} to obtain the evolved positions and velocities after the span of each age in the sample of 13,005 MSP ages described in Appendix~\ref{subap:age_distribution}. In this way, we compiled $10,000\times13,005=130$\,million evolved states (i.e., positions and velocities for different MSP ages) of simulated MSPs, from which $\sim4$\% evolved states outside the edge of the Galaxy (i.e., beyond 292\,kpc from the Galactic centre, \citealp{Deason20}) were discarded.   
To take into account the selection effect that astrometrically well determined pulsars are predominantly located within $\lesssim8$\,kpc, we removed evolved states that are beyond the 8\,kpc distance. Additionally, given that the observed MSPs compiled in this work are $\lesssim2.5$\,kpc from the Galactic plane (see Figure~\ref{fig:v_and_z_distributions}), we discarded the evolved states with $|z_\mathrm{e}|>2.5$\,kpc, where $z_\mathrm{e}$ refer to the evolved Galactic heights. The remaining evolved states are $\sim37$\,million in number, which amount to $\sim30$\% of the Galaxy-wide evolved states.

\subsection{Results of dynamical population synthesis}
\label{subsec:popSyn_results}

From the remaining evolved states, we calculated the evolved $v_\mathrm{l}$, $v_\mathrm{b}$, and $z_\mathrm{e}$ in the same way as the calculation of the observed counterparts (see Section~\ref{sec:obs} and references therein). 
We overlaid the results of the evolved $v_\mathrm{l}$, $v_\mathrm{b}$, and $z_\mathrm{e}$ to Figure~\ref{fig:v_and_z_distributions} in the same way as the initial counterparts.
As is shown in Figure~\ref{fig:v_and_z_distributions}, the initial $v_\mathrm{l}$ distribution (labeled as ``Tri-Normal'') is indeed in good consistency with the evolved $v_\mathrm{l}$ distribution, which validates the ``stable-$v_\mathrm{l}$'' assumption (see Section~\ref{sec:v_sys_distribution}) within the initial phase space of this study.
Despite the validation, we reiterate the necessity of assessing the consistency between the initial and the evolved $v_\mathrm{l}$ on a case-by-case basis, especially in the regime where $v_\mathrm{sys}$ and/or $\zeta_0$ are relatively large.
In contrast, the distributions of evolved $v_\mathrm{b}$ and $z_\mathrm{e}$ are considerably different from the initial counterparts, which meets our expectation that the initial MSP $z_\mathrm{i}$ would spread out over time, while the initial $v_\mathrm{b}$ becoming weakened due to the pull of the Galactic potential.

Moreover, from the Galaxy-wide $z_\mathrm{e}$ sample (of $\sim125$\,million), we found that the best-fit scale height $\zeta_1$ for field MSP systems in the whole Galaxy is 0.68\,kpc, which is 31\% larger than the observed scale height $\zeta^\mathrm{local}_\mathrm{obs}=0.52$\,kpc estimated from relatively nearby MSPs. 
Interestingly, unlike the observed $z$ sample that is best described by a Laplace distribution (see Appendix~\ref{subsubap:obs_z_PDF}), the Galaxy-wide $z_\mathrm{e}$ sample is best characterized by the Cauchy distribution with 0.48\,kpc scale (see Figure~\ref{fig:Galaxy_wide_Ze_distribution}).

In our analysis, $\zeta_0$ and $\zeta_1$ were estimated based on the observed $z$ and $v_\mathrm{b}$, without accounting for selection effects related to Galactic latitude (also see Appendix~\ref{subsubap:birth_scale_height}). 
As noted in Section~\ref{subsec:binary_vs_solitary}, 
the observed scale height of binary field MSPs is larger than solitary field MSPs. Therefore, the $\zeta_0$ and $\zeta_1$ estimated here should reflect the weighted average of the two sub-samples (of binary and solitary field MSPs). 
In the future, increasing MSP detections across the sky, driven by progressively advanced pulsar search techniques, will enable further refinement of $\zeta_0$ and $\zeta_1$.

\section{The indicative millisecond pulsar retention fractions in star clusters}
\label{sec:NS_retention_fractions}

At the time of writing, about 345 pulsars --- predominantly MSPs, have been found to be retained in 45 globular clusters (see the ``Pulsars in globular clusters'' Catalogue\footnote{\url{https://www3.mpifr-bonn.mpg.de/staff/pfreire/GCpsr.html}}). The high NS retention fraction of over 10\% to 20\% in globular clusters has been difficult to reconcile with the relatively large NS peculiar velocities 
\citep{Cordes97,Toscano99,Hobbs05,Gonzalez11} that are believed to easily exceed the generally small globular cluster central escape velocities ($\sim50$\,\kmps); this NS retention problem in globular clusters challenges the conventional NS formation theory \citep{Pfahl02}.
As the vast majority of pulsars detected in globular clusters are MSPs, the crux of the globular cluster NS retention problem is investigating why globular clusters host more MSPs than expected. 
However, the absence of the MSP $v_\mathrm{sys}$ distribution has led to significant uncertainties in the investigation.
Furthermore, while the mean 3D space velocities estimated for MSPs by \citet{Hobbs05,Gonzalez11} are consistent with our median $v_\mathrm{sys}$ (see Section~\ref{sec:v_sys_distribution}), the relatively large astrometric uncertainties in previous works would disproportionately inflate $v_\perp$ at the lower end of their $v_\perp$ distributions, which is expected to have lowered the predicted indicative rates of MSP retention in globular clusters.

Provided the null hypothesis that MSPs in star clusters follow the same formation channels (and accordingly the same $v_\mathrm{sys}$ distribution) as field MSP systems,
the $v_\mathrm{sys}$ CDF (see Equation~\ref{eq:Vsys_CDF}) derived for field MSP systems can also be seen as the indicative MSP retention fraction as a function of the star cluster central escape velocity $v_\mathrm{esc}$. At the nominal globular cluster $v_\mathrm{esc}=50$\,\kmps, we would expect $\approx14\%$ of MSPs to be retained in the globular cluster, which is no longer inconsistent with the expected retention fraction range of roughly 10\% to 20\% for binary systems \citep{Drukier96}.  
The indicative MSP retention fraction would further rise to $\gtrsim48$\% for a globular cluster with exceptionally high $v_\mathrm{esc}$ of $\gtrsim100$\,\kmps\ (e.g., Terzan~5).
Therefore, our MSP $v_\mathrm{sys}$ distribution may be able to account for the large number of MSPs detected in globular clusters, potentially alleviating concerns regarding NS retention in globular clusters. Accordingly, we see no evidence that the aforementioned null hypothesis (that MSPs in star clusters follow the same formation channels as field MSP systems) is violated. 
More accurate determination of the MSP retention fraction requires careful N-body simulations based on the MSP $v_\mathrm{sys}$ distribution, which is beyond the scope of this work.

\section{Summary and Prospects}
\label{sec:summary}

This work derives the systemic recoil velocity ($v_\mathrm{sys}$) distribution of field millisecond pulsar (MSP) systems from the observed $v_\mathrm{l}$ (the Galactic-longitude component of the transverse peculiar velocity), assuming that {\bf 1)} $v_\mathrm{l}$ is statistically stable over time (the ``stable-$v_\mathrm{l}$'' assumption), and {\bf 2)} the $\boldsymbol{v}_\mathrm{sys}$ directions are uniformly distributed (the ``isotropic-$\boldsymbol{v}_\mathrm{sys}$'' assumption). 
We found that the $v_\mathrm{sys}$ distribution can be approximated by a linear combination of three Maxwellian distributions.
Using dynamical population synthesis (DPS), we validated the ``stable-$v_\mathrm{l}$'' and evaluated the initial and current Galaxy-wide scale heights of field MSP systems by matching the observed Galactic heights and $v_\mathrm{l}$ (the Galactic-latitude component of the transverse peculiar velocity) to the DPS counterparts. The MSP $v_\mathrm{sys}$ distribution reported in this work predicts relatively large MSP retention fractions that may account for the large number of MSPs retained in globular clusters.

In light of the interesting discovery that the MSP $v_\mathrm{sys}$ distribution can be well approximated by a linear combination of three Maxwellian components, we highly encourage the production of template $v_\mathrm{sys}$ distribution for each formation channel using population synthesis that carefully simulates the MSP formation.
These template $v_\mathrm{sys}$ distributions can be used to compare with the observation-based $v_\mathrm{sys}$ distribution, thereby deepening the understanding of MSP formation.
Finally, the methods described in this paper can be used to determine the $v_\mathrm{sys}$ distribution and the scale height for other kinds of kinematically old compact stars (such as black hole systems) using precise astrometry (especially when radial velocity information is missing).

\begin{acknowledgements}
HD appreciates the helpful comments from the anonymous reviewer, and acknowledges the EACOA Fellowship awarded by the East Asia Core Observatories Association.
\end{acknowledgements}

%
%

\bibliographystyle{aa}
\bibliography{refs}

\begin{appendix} 

\section{Selection of the millisecond pulsar sample}
\label{ap:sample_selection}

We selected the MSP sample from the {\tt PSRCAT} Catalogue\textsuperscript{\ref{footnote:psrcat}} \citep{Manchester05} with the help of the {\tt psrqpy} Python package \citep{Pitkin18}. 
The MSPs were searched using the following criteria:
\begin{enumerate}[label=(\roman*)]
    \item $P_0<~$\mspplim, where $P_0$ represents the spin periods of MSPs, and
    \item $\dot{P}_0<2\times10^{-17}\,{\mathrm{s~s^{-1}}}$, where $\dot{P}_0$ denotes the time derivative of $P_0$, and
    \item MSPs should not belong to globular clusters (as explained in Section~\ref{sec:obs}), and
    \item both proper motion magnitude and parallax are of $>3\,\sigma$ significance, with the uncertainty in the proper motion magnitude calculated through the propagation of uncertainties in the right ascension (RA) and the declination components.
    
\end{enumerate}

There is no clear $P_0$ boundary between MSPs and other pulsars. Various $P_0$ upper limits up to $\sim100$\,ms have been adopted in different studies of MSPs \citep[e.g.][]{Manchester17,Spiewak22,Ding23}. In this work, we adopted the indicative $P_0$ upper limit of 60\,ms.
Additionally, we followed \citet{Spiewak22}, applying the filter $\dot{P}_0<2\times10^{-17}\,\mathrm{s~s^{-1}}$ to exclude recently born ($\lesssim10$\,kyr old) pulsars (such as the Crab pulsar with 34\,ms spin period), except for PSR~J1417$-$4402 that does not have a reported $\dot{P}_0$.
Despite the $\dot{P}_0$ being unknown to us, PSR~J1417$-$4402 is suggested to have entered the late recycling phase \citep{Camilo16a}, therefore likely being kinematically similar to many other MSPs.
Though meeting all the conditions, PSR~J1024$-$0719, which is in an unusually long-period orbit with a main-sequence star, is suggested to be born in (and ejected from) a globular cluster \citep{Kaplan16}. Therefore, we removed PSR~J1024$-$0719 from the MSP sample.

As mentioned in Section~\ref{sec:obs}, we look for field MSP systems with $>3\,\sigma$ parallax determinations. However, the Gaia-based astrometric results registered in the {\tt PSRCAT} Catalogue come from \citet{Jennings18} based on the Gaia Data Release (DR) 2 (instead of the latest Gaia DR3). Therefore, in practice, we conducted the sample search in three steps. 
In the first step, we searched for MSPs with a discounted parallax significance threshold of $3\times0.8=2.4$, where 0.8 is the expected ratio of parallax precision between the Gaia DR3 and the Gaia DR2.
In the second step, we updated the astrometric results of \citet{Jennings18} to Gaia DR3 utilizing the {\tt astroquery} package \citep{Ginsburg19}. In the final step, we removed MSPs with insignificant ($<3\,\sigma$) determinations of parallaxes or proper motions.
Compared to parallaxes, proper motions are usually determined at higher significance. The proper motion filter (as part of the criterion iv) only removes one MSP (i.e., PSR~J1811$-$2405) that is too close to the ecliptic plane to acquire precise proper motion in the declination direction with pulsar timing \citep{Ng20}.
Through the three-step procedure, we reached a sample of 64 field MSP systems, which are compiled in Table~\ref{tab:MSPs_d__v_t} along with their most precise proper motion and parallax determinations.

After calculating the transverse peculiar velocities $\boldsymbol{v}_\perp$ of the 64 MSPs (see Section~\ref{sec:obs}), we found no correlation between $P_0$ and $\boldsymbol{v}_\perp$, hence indicating negligible selection effect of $P_0$ on $\boldsymbol{v}_\perp$.

\subsection{High-precision samples}
\label{subap:obtain_high_precision_samples}
For the selection of the main sample, the thresholds for the significance of proper motions and parallaxes are set to be relatively low (i.e., $3\,\sigma$), as raising the thresholds would inevitably favour high-proper-motion and nearby pulsars. 
Higher proper motions would likely correspond to larger transverse peculiar velocities $\boldsymbol{v}_\perp$. 
On the other hand, given no indication of correlation between $\boldsymbol{v}_\perp$ and MSP distances, raising the threshold of parallax significance could be rewarding: 
further lowering astrometric uncertainties, which are dominated by the distance uncertainties, might hold the key to potentially resolving multiple velocity modes.
In order to investigate the impact of fractional parallax precision on the velocity analysis, we also adopted higher thresholds of parallax significance for sample selection. For example, with the $5\,\sigma$ and the $7\,\sigma$ thresholds of parallax significances, two ``high-precision'' samples of 45 and 30 MSPs, respectively, were obtained for comparison with the full sample (of 64 MSPs) (see Table~\ref{tab:diptests}).

\begin{figure*}[h]
    \centering    \includegraphics[width=0.98\textwidth]{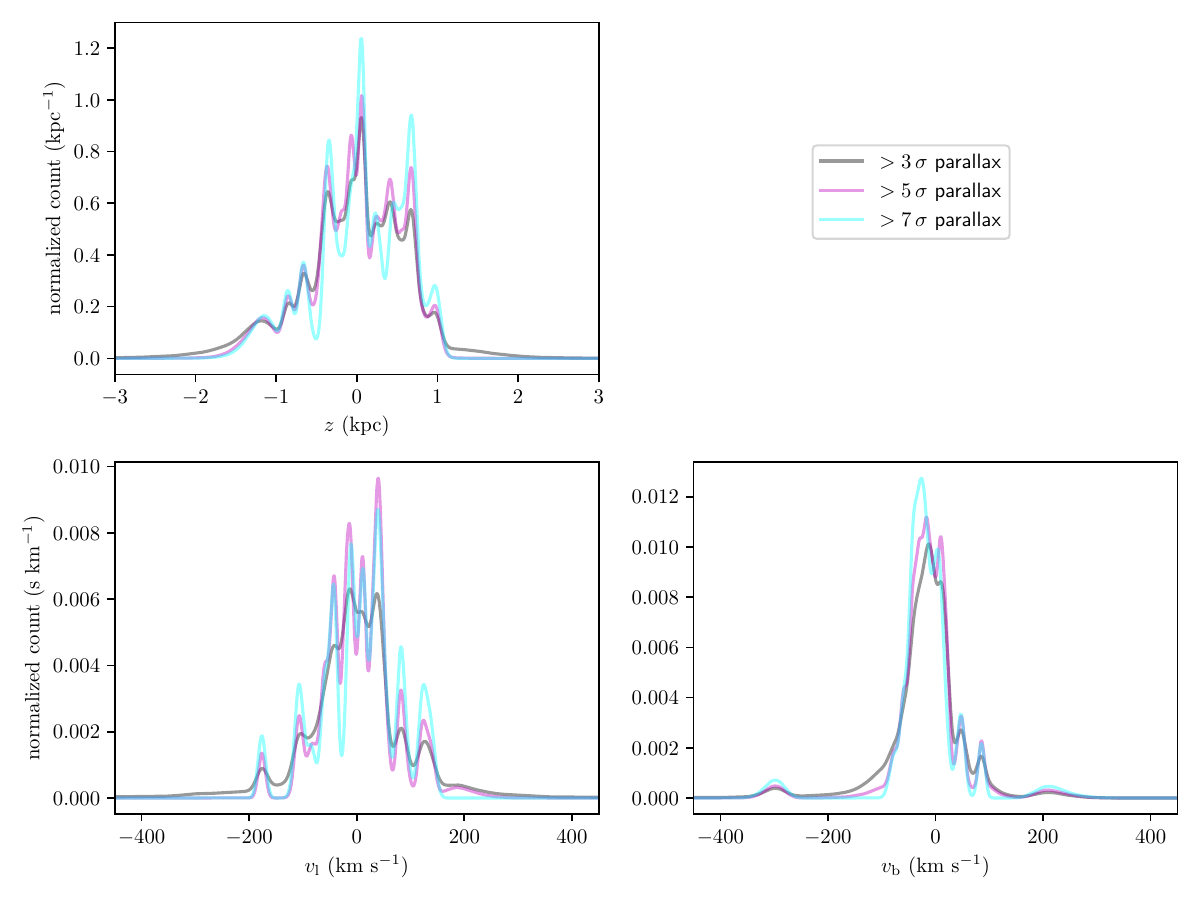}
    \caption{
    The upper-left, lower-left and lower-right panels display the normalized histograms of Galactic heights $z$, the Galactic-longitude component $v_\mathrm{l}$ and the Galactic-latitude component $v_\mathrm{b}$ of transverse peculiar velocities, respectively. 
    The normalized histograms have been smoothed with kernel density estimation (using {\tt scipy.stats.Gaussian\_kde}).
    Each histogram is concatenated from 10000 simulations drawn from the assumed split normal distributions for the measurements (of $z$, $v_\mathrm{l}$ or $v_\mathrm{b}$; see Table~\ref{tab:MSPs_d__v_t}).  The black, magenta and yellow curves correspond to the full sample (of 64 field MSPs) and two high-precision samples (see Appendix~\ref{subap:obtain_high_precision_samples}), respectively.}
    \label{fig:v_and_z_distributions_for_high_precision_samples}
\end{figure*}

In the same way as described in Section~\ref{sec:obs}, we compiled $v_\mathrm{l}$, $v_\mathrm{b}$ and $z$ chains for the high-precision MSP samples, which are displayed in Figure~\ref{fig:v_and_z_distributions_for_high_precision_samples} along with the counterparts of the full sample.
Compared to the full sample of 64 MSPs, the high-precision MSP samples render consistent $v_\mathrm{l}$, $v_\mathrm{b}$, and $z$ distributions, while showing increasingly prominent substructures in the $v_\mathrm{l}$, $v_\mathrm{b}$, and $z$ distributions as sample size drops, likely due to small sample effects.

\begin{table*}[h]
     \centering
     \caption[]{\label{tab:MSPs_d__v_t}
     Distances $d$, Galactic heights $z$, transverse peculiar velocities $\textit{\textbf{v}}_\perp\equiv\left(v_\mathrm{l}, v_\mathrm{b}\right)$, and their magnitudes $v_\perp$ of the 64 selected millisecond pulsars calculated from their published proper motions ($\mu_\alpha$ and $\mu_\delta$) and parallaxes $\varpi$, where {\bf 1)} negative $z$ values correspond to negative Galactic latitudes $b$, and {\bf 2)} $v_\mathrm{b}$ and $v_\mathrm{l}$ represent the transverse peculiar velocity components along the directions of $b$ and Galactic longitudes $l$, respectively.}
     {\renewcommand{\arraystretch}{1.3}
     \resizebox{0.95\textwidth}{!}{
    \begin{tabular}{ccccccccccc}
        \hline 
        \hline
        PSR & solitary? & $\mu_\alpha$ & $\mu_\delta$  &  $\varpi$ & Ref. \tablefootmark{a} & $d$
        &  $z$ & $v_l$ & $v_b$ & $v_\perp$ \\
        & (Y/N) & (\maspy) & (\maspy) & (mas) & & (kpc) & (kpc) & (\kmps) & (\kmps) & (\kmps) \\ 
        \hline
        
J0023$+$0923 & N & $-11.00(7)$ & $-8.8(1)$ & $0.6(1)$ & (1), (2) & $1.9^{+0.5}_{-0.3}$ & $-1.5^{+0.3}_{-0.4}$ & $-88^{+16}_{-24}$ & $-41^{+9}_{-13}$ & $98^{+28}_{-19}$ \\ 
J0030$+$0451 & Y & $-6.15(5)$ & $0.4(1)$ & $3.04(5)$ & (3) & $0.329(5)$ & $-0.278(5)$ & $-9.5(1)$ & $12.2(2)$ & $15.4(2)$ \\ 
J0125$-$2327 & N & $37.14(8)$ & $10.5(2)$ & $0.8(1)$ & (4), (5) & $1.2^{+0.2}_{-0.1}$ & $-1.2^{+0.1}_{-0.2}$ & $35^{+7}_{-5}$ & $216^{+34}_{-25}$ & $220^{+34}_{-26}$ \\ 
J0437$-$4715 & N & $121.4420(6)$ & $-71.4717(7)$ & $6.43(5)$ & (6) & $0.156(1)$ & $-0.1040(8)$ & $46.3(4)$ & $85.3(7)$ & $97.0(8)$ \\ 
J0610$-$2100 & N & $9.11(3)$ & $16.45(3)$ & $0.7(1)$ & (3), (7) & $1.5^{+0.3}_{-0.2}$ & $-0.46^{+0.07}_{-0.09}$ & $-72^{+10}_{-14}$ & $95^{+20}_{-14}$ & $119^{+23}_{-17}$ \\ 
J0613$-$0200 & N & $1.837(2)$ & $-10.359(6)$ & $1.00(5)$ & (8) & $1.01(5)$ & $-0.163^{+0.008}_{-0.009}$ & $44(3)$ & $-17.7^{+0.8}_{-0.9}$ & $47(3)$ \\ 
J0614$-$3329 & N & $0.58(9)$ & $-1.9(1)$ & $1.5(4)$ & (4), (9) & $1.0^{+0.7}_{-0.3}$ & $-0.4^{+0.1}_{-0.3}$ & $21^{+18}_{-8}$ & $-8^{+1}_{-4}$ & $23^{+18}_{-8}$ \\ 
J0621$+$1002 & N & $3.27(9)$ & $-1.1(3)$ & $0.7(1)$ & (3) & $1.6^{+0.5}_{-0.3}$ & $-0.06^{+0.01}_{-0.02}$ & $10^{+6}_{-4}$ & $17^{+5}_{-3}$ & $20^{+8}_{-5}$ \\ 
J0636$+$5128 & N & $1.1(4)$ & $-4.4(7)$ & $1.4(2)$ & (2), (10) & $0.8^{+0.2}_{-0.1}$ & $0.26^{+0.06}_{-0.04}$ & $7^{+5}_{-4}$ & $-5(2)$ & $9^{+5}_{-3}$ \\ 
J0636$-$3044 & Y & $26.01(9)$ & $-19.3(1)$ & $4.3(7)$ & (4) & $0.26^{+0.07}_{-0.04}$ & $-0.07^{+0.01}_{-0.02}$ & $34^{+10}_{-6}$ & $17^{+5}_{-3}$ & $38^{+11}_{-7}$ \\ 
J0645$+$5158 & Y & $1.55(2)$ & $-7.47(4)$ & $0.8(2)$ & (2) & $1.4^{+0.4}_{-0.3}$ & $0.47^{+0.13}_{-0.09}$ & $39^{+14}_{-9}$ & $-10(2)$ & $40^{+14}_{-9}$ \\ 
J0737$-$3039A & N & $-2.57(3)$ & $2.08(4)$ & $1.3(1)$ & (11) & $0.80^{+0.09}_{-0.08}$ & $-0.063^{+0.006}_{-0.007}$ & $1.4^{+0.6}_{-0.5}$ & $-6.2^{+0.5}_{-0.7}$ & $6.4^{+0.8}_{-0.6}$ \\ 
J0751$+$1807 & N & $-2.70(1)$ & $-13.27(7)$ & $0.85(4)$ & (8) & $1.18^{+0.06}_{-0.05}$ & $0.43(2)$ & $55(3)$ & $-37(2)$ & $67^{+4}_{-3}$ \\ 
J1012$+$5307 & N & $2.624(3)$ & $-25.487(4)$ & $1.14(4)$ & (8), (3) & $0.88(3)$ & $0.68(2)$ & $82(3)$ & $49(2)$ & $95(4)$ \\ 
J1022$+$1001 & N & $-14.92(5)$ & $5.61(4)$ & $1.39(4)$ & (12) & $0.72(2)$ & $0.56(2)$ & $-42.2^{+1.0}_{-1.1}$ & $-21.8^{+0.8}_{-0.9}$ & $47(1)$ \\ 
J1023$+$0038 & N & $4.76(3)$ & $-17.34(4)$ & $0.73(3)$ & (13) & $1.37^{+0.06}_{-0.05}$ & $0.98(4)$ & $122(5)$ & $-33(2)$ & $126^{+6}_{-5}$ \\ 
J1125$-$6014 & N & $11.09(3)$ & $-13.00(3)$ & $1.2(3)$ & (4) & $2^{+5}_{-1}$ & $0.03^{+0.08}_{-0.02}$ & $210^{+484}_{-108}$ & $-87^{+45}_{-203}$ & $234^{+536}_{-123}$ \\ 
J1227$-$4853 & N & $-18.8(1)$ & $7.30(9)$ & $0.5(1)$ & (14) & $2.9^{+1.5}_{-0.8}$ & $0.7^{+0.4}_{-0.2}$ & $-183^{+50}_{-91}$ & $71^{+39}_{-20}$ & $195^{+97}_{-52}$ \\ 
J1300$+$1240 & N & $45.50(4)$ & $-84.70(7)$ & $1.41(8)$ & (16), (15) & $0.72(4)$ & $0.69(4)$ & $125^{+7}_{-6}$ & $-300^{+17}_{-19}$ & $325^{+20}_{-18}$ \\ 
J1400$-$1431 & N & $17(2)$ & $-55(6)$ & $4(1)$ & (17) & $0.7^{+1.2}_{-0.4}$ & $0.5^{+0.9}_{-0.3}$ & $-2^{+9}_{-22}$ & $-201^{+105}_{-341}$ & $201^{+348}_{-107}$ \\ 
J1417$-$4402 & N & $-4.76(4)$ & $-5.10(5)$ & $0.20(5)$ & (14) & $5.3^{+1.6}_{-1.0}$ & $1.5^{+0.4}_{-0.3}$ & $-14^{+18}_{-9}$ & $-100^{+18}_{-23}$ & $101^{+22}_{-16}$ \\ 
J1431$-$4715 & N & $-11.8(1)$ & $-14.5(2)$ & $0.5(1)$ & (14) & $2.6^{+1.4}_{-0.7}$ & $0.5^{+0.3}_{-0.1}$ & $-149^{+38}_{-67}$ & $-114^{+31}_{-62}$ & $188^{+91}_{-49}$ \\ 
J1446$-$4701 & N & $-4.24(9)$ & $-2.4(2)$ & $0.7(2)$ & (4) & $3^{+3}_{-1}$ & $0.5^{+0.5}_{-0.2}$ & $-8^{+22}_{-3}$ & $-12^{+5}_{-10}$ & $15^{+14}_{-6}$ \\ 
J1455$-$3330 & N & $7.85(1)$ & $-1.98(4)$ & $1.3(1)$ & (8) & $0.79^{+0.07}_{-0.06}$ & $0.30^{+0.03}_{-0.02}$ & $36(2)$ & $-23(2)$ & $43(3)$ \\ 
J1518$+$4904 & N & $-0.68(3)$ & $-8.53(4)$ & $1.24(4)$ & (3) & $0.81^{+0.03}_{-0.02}$ & $0.66(2)$ & $-13.5(5)$ & $8.6^{+0.6}_{-0.5}$ & $16.0^{+0.8}_{-0.7}$ \\ 
J1537$+$1155 & N & $1.484(7)$ & $-25.29(1)$ & $1.07(7)$ & (3) & $0.95^{+0.07}_{-0.06}$ & $0.71^{+0.05}_{-0.04}$ & $-85^{+6}_{-7}$ & $-56^{+3}_{-4}$ & $102^{+8}_{-7}$ \\ 
J1545$-$4550 & N & $-0.51(7)$ & $2.5(1)$ & $0.4(1)$ & (4), (18) & $5^{+4}_{-2}$ & $0.6^{+0.4}_{-0.2}$ & $112^{+197}_{-60}$ & $40^{+34}_{-16}$ & $120^{+202}_{-61}$ \\ 
J1600$-$3053 & N & $-0.943(3)$ & $-6.92(1)$ & $0.72(2)$ & (8) & $1.39(4)$ & $0.39(1)$ & $-22.5^{+0.8}_{-0.9}$ & $-31.8(9)$ & $39(1)$ \\ 
J1614$-$2230 & N & $3.9(1)$ & $-32.1(5)$ & $1.5(1)$ & (2) & $0.66^{+0.05}_{-0.04}$ & $0.23^{+0.02}_{-0.01}$ & $-55^{+4}_{-5}$ & $-78^{+5}_{-6}$ & $95(7)$ \\ 
J1629$-$6902 & Y & $-6.50(3)$ & $-8.44(4)$ & $0.9(2)$ & (4) & $1.5^{+0.8}_{-0.4}$ & $-0.36^{+0.09}_{-0.20}$ & $-46^{+13}_{-25}$ & $-1.6(3)$ & $45^{+25}_{-12}$ \\ 
J1640$+$2224 & N & $2.102(4)$ & $-11.333(7)$ & $0.73(6)$ & (8), (3) & $1.4(1)$ & $0.86^{+0.08}_{-0.07}$ & $-44(4)$ & $-30(2)$ & $53^{+5}_{-4}$ \\ 
J1643$-$1224 & N & $6.1(4)$ & $3(2)$ & $1.1(1)$ & (3) & $0.94^{+0.10}_{-0.08}$ & $0.34^{+0.04}_{-0.03}$ & $38^{+8}_{-7}$ & $-13(5)$ & $40^{+6}_{-5}$ \\ 
J1658$-$5324 & Y & $0.0(1)$ & $2.9(2)$ & $1.3(3)$ & (4) & $1.4^{+3.2}_{-0.5}$ & $-0.16^{+0.06}_{-0.37}$ & $33^{+95}_{-9}$ & $13^{+33}_{-5}$ & $35^{+105}_{-11}$ \\ 
J1713$+$0747 & N & $4.9215(8)$ & $-3.920(2)$ & $0.95(6)$ & (8), (19) & $1.07^{+0.07}_{-0.06}$ & $0.45(3)$ & $9.8(1)$ & $-28(2)$ & $29^{+2}_{-1}$ \\ 
J1723$-$2837 & N & $-11.73(4)$ & $-24.05(3)$ & $1.07(4)$ & (14) & $0.94(4)$ & $0.070(3)$ & $-109(5)$ & $-17.1^{+0.7}_{-0.8}$ & $110(5)$ \\ 
J1730$-$2304 & Y & $20.1(1)$ & $-4.7(6)$ & $2.0(1)$ & (3) & $0.51^{+0.03}_{-0.02}$ & $0.053(3)$ & $27^{+2}_{-1}$ & $-46^{+2}_{-3}$ & $54^{+3}_{-2}$ \\ 
J1738$+$0333 & N & $7.09(1)$ & $5.07(3)$ & $0.59(5)$ & (8), (3) & $1.7^{+0.2}_{-0.1}$ & $0.53^{+0.05}_{-0.04}$ & $87^{+8}_{-7}$ & $-27(2)$ & $91^{+8}_{-7}$ \\ 
J1741$+$1351 & N & $-8.97(1)$ & $-7.44(2)$ & $0.6(1)$ & (2), (20) & $1.8^{+0.4}_{-0.3}$ & $0.7^{+0.2}_{-0.1}$ & $-59^{+10}_{-13}$ & $53^{+13}_{-9}$ & $79^{+19}_{-13}$ \\ 
J1744$-$1134 & Y & $18.806(2)$ & $-9.39(1)$ & $2.58(3)$ & (8) & $0.388(5)$ & $0.0618(7)$ & $13.43(4)$ & $-38.5(4)$ & $40.8(4)$ \\ 
J1757$-$5322 & N & $-2.48(8)$ & $-10.0(2)$ & $1.2(3)$ & (4) & $1.4^{+1.9}_{-0.5}$ & $-0.3^{+0.1}_{-0.4}$ & $-53^{+22}_{-67}$ & $-13^{+4}_{-14}$ & $54^{+72}_{-23}$ \\ 
J1804$-$2717 & N & $2.46(2)$ & $-16.9(4)$ & $1.1(3)$ & (8) & $6^{+7}_{-4}$ & $-0.3^{+0.2}_{-0.3}$ & $-251^{+148}_{-91}$ & $-279^{+191}_{-348}$ & $412^{+307}_{-276}$ \\ 
J1832$-$0836 & Y & $-8.06(5)$ & $-21.0(2)$ & $0.5(1)$ & (2), (18) & $6^{+5}_{-3}$ & $0.03^{+0.02}_{-0.01}$ & $-493^{+214}_{-269}$ & $-71^{+35}_{-62}$ & $497^{+289}_{-217}$ \\ 
J1853$+$1303 & N & $-1.62(2)$ & $-2.96(4)$ & $0.53(7)$ & (3) & $2.0^{+0.4}_{-0.3}$ & $0.19^{+0.03}_{-0.02}$ & $13^{+3}_{-2}$ & $3.5^{+0.8}_{-0.6}$ & $14^{+3}_{-2}$ \\ 
J1857$+$0943 & N & $-2.670(3)$ & $-5.428(6)$ & $0.89(6)$ & (8) & $1.15^{+0.09}_{-0.07}$ & $0.061^{+0.005}_{-0.004}$ & $-7.7^{+0.8}_{-0.9}$ & $0.11^{+0.05}_{-0.04}$ & $7.7^{+0.9}_{-0.8}$ \\ 
J1909$-$3744 & N & $-9.523(1)$ & $-35.780(5)$ & $0.94(2)$ & (8) & $1.07(2)$ & $-0.357^{+0.007}_{-0.008}$ & $-177(4)$ & $-15.5(3)$ & $177(4)$ \\ 
J1910$+$1256 & N & $0.24(2)$ & $-7.10(3)$ & $0.36(6)$ & (8), (3) & $3.2^{+0.9}_{-0.6}$ & $0.10^{+0.03}_{-0.02}$ & $-12.7^{+4.2}_{-0.9}$ & $-52^{+9}_{-14}$ & $54^{+13}_{-8}$ \\ 
J1911$+$1347 & Y & $-2.900(5)$ & $-3.684(9)$ & $0.40(9)$ & (8) & $3.5^{+2.1}_{-1.0}$ & $0.11^{+0.07}_{-0.03}$ & $16^{+33}_{-9}$ & $16^{+9}_{-4}$ & $23^{+32}_{-9}$ \\ 
J1915$+$1606 & N & $-0.7(1)$ & $-0.0(1)$ & $0.24(7)$ & (21) & $6^{+3}_{-2}$ & $0.23^{+0.11}_{-0.07}$ & $184^{+86}_{-66}$ & $20^{+9}_{-6}$ & $183^{+86}_{-64}$ \\ 
J1918$-$0642 & N & $-7.124(5)$ & $-5.96(2)$ & $0.71(7)$ & (8), (3) & $1.5^{+0.2}_{-0.1}$ & $-0.23^{+0.02}_{-0.03}$ & $-37^{+4}_{-5}$ & $23^{+3}_{-2}$ & $44^{+5}_{-4}$ \\ 
J1933$-$6211 & N & $-5.62(1)$ & $11.09(3)$ & $0.7(2)$ & (4), (22) & $1.9^{+1.1}_{-0.5}$ & $-0.9^{+0.3}_{-0.5}$ & $118^{+68}_{-31}$ & $68^{+42}_{-20}$ & $136^{+79}_{-37}$ \\ 
J1939$+$2134 & Y & $0.074(2)$ & $-0.410(3)$ & $0.35(3)$ & (3) & $3.0^{+0.3}_{-0.2}$ & $-0.015(1)$ & $80^{+9}_{-7}$ & $-3.8^{+0.3}_{-0.4}$ & $80^{+9}_{-8}$ \\

        
    \end{tabular}
    }}
    
    \end{table*}

\begin{table*}[h]
     \centering
     \contcaption{\textit{continued}}
     {\renewcommand{\arraystretch}{1.3}
     \resizebox{0.95\textwidth}{!}{
    \begin{tabular}{ccccccccccc}
        \hline 
        \hline
        PSR  & solitary? & $\mu_\alpha$ & $\mu_\delta$  &  $\varpi$ & Ref. \tablefootmark{a} & $d$
        &  $z$ & $v_l$ & $v_b$ & $v_\perp$ \\
         & (Y/N) & (\maspy) & (\maspy) & (mas) & & (kpc) & (kpc) & (\kmps) & (\kmps) & (\kmps)  \\ 
        \hline

J1946$-$5403 & N & $-1.08(3)$ & $-4.75(5)$ & $0.8(1)$ & (4) & $1.3(2)$ & $-0.65^{+0.08}_{-0.11}$ & $-17^{+3}_{-5}$ & $7^{+2}_{-1}$ & $19^{+5}_{-4}$ \\ 
J2043$+$1711 & N & $-5.70(1)$ & $-10.84(2)$ & $0.72(6)$ & (2) & $1.4(1)$ & $-0.38^{+0.03}_{-0.04}$ & $-42(4)$ & $-13(1)$ & $44^{+5}_{-4}$ \\ 
J2124$-$3358 & Y & $-14.09(1)$ & $-50.32(3)$ & $2.1(1)$ & (8) & $0.48(2)$ & $-0.34(2)$ & $-107(6)$ & $20.6^{+1.0}_{-0.9}$ & $109(6)$ \\ 
J2129$-$0429 & N & $12.10(7)$ & $10.19(6)$ & $0.48(7)$ & (1), (14) & $2.1^{+0.4}_{-0.3}$ & $-1.3(2)$ & $194^{+33}_{-25}$ & $-63^{+9}_{-12}$ & $204^{+35}_{-27}$ \\ 
J2145$-$0750 & N & $-9.7(3)$ & $-8.2(9)$ & $1.60(6)$ & (12) & $0.63(2)$ & $-0.42(2)$ & $-21(3)$ & $14(2)$ & $25(2)$ \\ 
J2222$-$0137 & N & $44.707(5)$ & $-5.40(1)$ & $3.72(1)$ & (23) & $0.269(1)$ & $-0.1935(7)$ & $39.8(1)$ & $-43.3(2)$ & $58.8(2)$ \\ 
J2234$+$0611 & N & $25.30(2)$ & $9.71(5)$ & $1.03(4)$ & (24) & $0.97(4)$ & $-0.66(3)$ & $139(5)$ & $-44(2)$ & $145^{+6}_{-5}$ \\ 
J2234$+$0944 & N & $6.9(2)$ & $-33.2(5)$ & $1.4(3)$ & (4), (25) & $0.8^{+0.2}_{-0.1}$ & $-0.52^{+0.10}_{-0.15}$ & $-43^{+9}_{-13}$ & $-108^{+21}_{-33}$ & $117^{+36}_{-23}$ \\ 
J2241$-$5236 & N & $18.881(4)$ & $-5.294(5)$ & $0.96(4)$ & (18) & $1.05^{+0.05}_{-0.04}$ & $-0.86^{+0.03}_{-0.04}$ & $-62(3)$ & $-61(2)$ & $87(4)$ \\ 
J2256$-$1024 & N & $3(1)$ & $-8(2)$ & $0.5(2)$ & (26) & $2.1^{+0.7}_{-0.5}$ & $-1.8^{+0.4}_{-0.6}$ & $-18^{+18}_{-20}$ & $-79^{+23}_{-33}$ & $83^{+36}_{-25}$ \\ 
J2317$+$1439 & N & $-1.46(10)$ & $3.7(2)$ & $0.60(8)$ & (2) & $1.7^{+0.3}_{-0.2}$ & $-1.2^{+0.1}_{-0.2}$ & $40^{+6}_{-5}$ & $42^{+6}_{-5}$ & $58^{+9}_{-7}$ \\ 
J2322$+$2057 & Y & $-18.30(5)$ & $-14.9(1)$ & $1.2(3)$ & (4), (8) & $1.0^{+0.5}_{-0.2}$ & $-0.6^{+0.2}_{-0.3}$ & $-91^{+22}_{-42}$ & $-19^{+6}_{-11}$ & $93^{+43}_{-23}$ \\ 
J2322$-$2650 & N & $-2.37(9)$ & $-8.2(2)$ & $1.3(2)$ & (4) & $0.8^{+0.2}_{-0.1}$ & $-0.8^{+0.1}_{-0.2}$ & $-22^{+4}_{-6}$ & $7.4^{+0.6}_{-0.5}$ & $23^{+6}_{-4}$ \\

        \hline
        
        \hline
    \end{tabular}
    }}
    \tablefoot{
    When an MSP has multiple proper motions and parallaxes reported by different studies, the values with smaller uncertainties (instead of higher significances) are adopted. 
    Despite our meticulous efforts, some of the compiled proper motions and parallaxes may not represent the most precise measurements available. 
    \\    
    \tablefoottext{a}{References for proper motions and parallaxes: (1) \citet{Bangale24}, (2) \citet{alam20}, (3) \citet{Ding23}, (4) \citet{Shamohammadi24}, (5) \citet{curyo23}, (6) \citet{reardon24}, (7) \citet{van-der-Wateren22}, (8) \citet{Antoniadis23}, (9) \citet{Guillemot16}, (10) \citet{fiore23}, (11) \citet{Kramer21a}, (12) \citet{Deller19}, (13) \citet{deller12}, (14) \citet{Jennings18}, (15) \citet{Konacki03}, (16) \citet{Yan13}, (17) \citet{Swiggum17}, (18) \citet{reardon21}, (19) \citet{Chatterjee09}, (20) \citet{Arzoumanian18a}, (21) \citet{Deller18}, (22) \citet{geyer23}, (23) \citet{ding24c}, (24) \citet{Stovall19}, (25) \citet{alam21}, and (26) \citet{Crowter20}. \\}
    
    }
    \end{table*}

\section{Determining the probability density function for $v_\mathrm{l}$}
\label{ap:determining_vl_PDF}

\subsection{Candidate probability density functions of $v_\mathrm{l}$}
\label{subap:candidate_vl_PDFs}

Following the first and the second steps described in Section~\ref{sec:v_sys_distribution}, we identified two prime groups of candidate probability density functions (PDFs) of $v_\mathrm{l}$. The first group is a linear combination of Cauchy distributions where the probability density $p$ takes the form of 
\begin{equation}
\label{eq:multi_scale_cauchy_PDF}
\begin{cases}
    p = \displaystyle\sum_{i}^{N} \frac{\alpha_i}{\pi\gamma_i \left(1+ \frac{{v_\mathrm{l}}^2}{{\gamma_i}^2}\right)}\,\,\,\,\,\,\,\,\,\,(\alpha_i>0 \text{ and } \gamma_i>0)\\[16pt]
    \displaystyle\sum_{i}^{N} \alpha_i = 1\,\,\,,
\end{cases}
\end{equation}
while the second group being a linear combination of normal distributions, following the relation
\begin{equation}
\label{eq:multi_scale_normal_PDF}
\begin{cases}
    p = \displaystyle\sum_{i}^{N} \frac{\alpha_i}{\sqrt{2\pi{\gamma_i}^2}}\exp{\left(-\frac{{v_\mathrm{l}}^2}{2{\gamma_i}^2}\right)}\,\,\,\,\,\,\,\,\,\,(\alpha_i>0)\\[16pt]
    \displaystyle\sum_{i}^{N} \alpha_i = 1\,\,\,.
\end{cases}
\end{equation}
In this paper, we refer to Equations~\ref{eq:multi_scale_cauchy_PDF} and \ref{eq:multi_scale_normal_PDF} as Cauchy series and Gaussian series, respectively.
In either equation, $\gamma_i$ is the scale of the $i$-th Cauchy or normal distribution.
When $N=1$, Equations~\ref{eq:multi_scale_cauchy_PDF} and \ref{eq:multi_scale_normal_PDF} are simplified to a Cauchy distribution and a normal distribution, respectively. For both groups, we refer to a PDF with $N=2,3,4$ as a two-component distribution, a three-component distribution, and a four-component distribution, respectively.
The best-fit parameters for Equations~\ref{eq:multi_scale_cauchy_PDF} and \ref{eq:multi_scale_normal_PDF} are provided in Table~\ref{tab:PDF_candidates}.

\begin{table*}[h]
     \centering
     \caption[]{\label{tab:PDF_candidates}
     Best-fit parameters of candidate $v_\mathrm{l}$ probability density functions defined by Equations~\ref{eq:multi_scale_cauchy_PDF} and \ref{eq:multi_scale_normal_PDF}, along with the Bayes factors $K$ and the goodness differences $\Delta\Lambda$ calculated with the best-fit Cauchy distribution ($N=1$ and $\gamma_1=51.2$\,\kmps) as the reference of comparison.}
     {\renewcommand{\arraystretch}{1.3}
     \resizebox{0.8\textwidth}{!}{
    \begin{tabular}{c|cccccccc|cc}
        \hline 
        \hline
        $N$ & $\gamma_1$ & $\alpha_1$ & $\gamma_2$ & $\alpha_2$ & $\gamma_3$ & $\alpha_3$ & $\gamma_4$ & $\alpha_4$ & $\ln{K}$ & $\Delta\Lambda$ \\
         & (\kmps) & & (\kmps) & & (\kmps) &  & (\kmps) & & & \\
          
        \hline
        \multicolumn{11}{c}{Cauchy series (Eq.~\ref{eq:multi_scale_cauchy_PDF})}\\
        \hline
        1 & 51.2 & 100\% & -- & -- & -- & -- & -- & --  & 0 & 0 \\
        2 & 51.2 & 99.7\% & 52.8 & 0.3\% & -- & -- & -- & -- & 0 & $-2$ \\
        \hline
        \multicolumn{11}{c}{Gaussian series (Eq.~\ref{eq:multi_scale_normal_PDF})}\\
        \hline
        1 & 68.3 & 100\% & -- & -- & -- & -- & --  & -- & $-13^{+8}_{-12}$ & $-13^{+8}_{-12}$ \\
        2 & 44.4 & 50.7\% & 123.9 & 49.8\% & -- & -- & -- & -- & $-2^{+2}_{-4}$ & $-4^{+2}_{-4}$ \\
        3 & 43.1 & 46.3\% & 107.6 & 47.7\% & 406.1 & 6.0\% & -- & -- & 0.6(2) & $-3.4(2)$ \\
        4 & 43.1 & 46.3\% & 107.3 & 47.6\% & 371.6 & 5.4\% & 776.1 & 0.7\% & 0.6(2) & $-5.4(2)$ \\
        \hline

        \hline
        
    \end{tabular}
    }}
    \tablefoot{
    Despite our exhaustive efforts, the best-fit two-component Cauchy distribution closely approaches the one-component Cauchy distribution, as the scales of the two components are nearly identical.   
    }
    \end{table*}

\subsection{Selecting from the candidate probability density distributions}
\label{subap:Bayes_analysis}

From each of the two prime PDF groups introduced in Appendix~\ref{subap:candidate_vl_PDFs}, we selected one $v_\mathrm{l}$ PDF based on a criterion that is equivalent to the Akaike information criterion \citep{Akaike11}. Specifically, the goodness $\Lambda$ of a PDF is quantified by
\begin{equation}
\label{eq:PDF_goodness}
    \Lambda = \ln{L}-k\,\,,    
\end{equation}
where $L$ refers to the likelihood that the measurements match the PDF, and $k$ represents the number of free parameters in the PDF.
A PDF with higher $\Lambda$ is considered better.
For Equations~\ref{eq:multi_scale_cauchy_PDF} and \ref{eq:multi_scale_normal_PDF},
\begin{equation}
\label{eq:k_and_N_sc}
k=2N-1\,\,;
\end{equation}
the likelihood $L$ can be calculated by
\begin{equation}
\label{eq:PDF_likelihood}
L = \displaystyle\prod_{j=1}^{64} p\left(v^j_\mathrm{l}\right)\,\,,
\end{equation}
where $p\left(v^j_\mathrm{l}\right)$ is the value of Equation~\ref{eq:multi_scale_cauchy_PDF} or \ref{eq:multi_scale_normal_PDF} at the observed $v^j_\mathrm{l}\,\,(j=1,2,...,64)$.

Accordingly, the comparison of two PDFs is based on the $\Lambda$ difference
\begin{equation}
\label{eq:goodness_difference}
    \Delta\Lambda=\Lambda_1-\Lambda_2 =\ln{\left(\frac{L_1}{L_2}\right)} - (k_1-k_2) = \ln{K} - 2\left( N_1-N_2\right)\,\,,
\end{equation}
where $K\equiv L_1/L_2$ is the Bayes factor between the two PDFs.
To evaluate the uncertainty in $K$ due to the uncertainties in the 64 $v_\mathrm{l}$ measurements, we assumed the observed $v_\mathrm{l}$ follow split normal distributions, and simulated 10000 sets of 64 $v_\mathrm{l}$ values. From each set of 64 $v_\mathrm{l}$ values we derived a $K$ based on Equation~\ref{eq:PDF_likelihood}. With the altogether 10000 derived $K$ values, we estimated $K$ and its uncertainty for two PDFs of comparison. The calculated $\ln{K}$ (as well as their uncertainties) of the candidate PDFs with respect to the best-fit Cauchy distribution are provided in Table~\ref{tab:PDF_candidates}.

With larger $N$ (and accordingly more free parameters), the PDF is expected to better approach the data, leading to larger $K$. Meanwhile, the incremental increase of $\ln{K}$ would quickly drop at larger $N$.
For each of the two prime groups of candidate PDFs, we started fitting the PDF parameter(s) and estimating the corresponding $K$ from $N=1$, then repeated the parameter fitting and $K$ estimation at incrementally larger $N$ until the incremental increase of $\ln{K}$ becomes negligible compared to 2 -- the incremental increase of the number of free parameters.
We found that $\Lambda$ (defined by Equations~\ref{eq:PDF_goodness} through \ref{eq:PDF_likelihood}) peaks at $N=1$ for the Cauchy series, and at $N=3$ for the Gaussian series (see Table~\ref{tab:PDF_candidates}). In other words, the best-fit Cauchy distribution and the best-fit three-component normal distribution are selected for the two prime groups of candidate PDFs.

Between the two finalist PDFs, the Cauchy distribution is preferred over the three-component normal distribution according to the $\Lambda$ standard (see Table~\ref{tab:PDF_candidates}). 
Therefore, we conclude that the sample of 64 observed $v_\mathrm{l}$ is well described by the simple PDF
\begin{equation}
\label{eq:vl_distribution}
\frac{p}{1\,\mathrm{s~{km}^{-1}}}=\frac{1}{51.2\pi\left[1+\left(\frac{{v_\mathrm{l}}}{51.2\,\mathrm{km~s^{-1}}}\right)^{2}\right]}\,\,.
\end{equation}
On the other hand, the three-component normal distribution is highly convenient for the calculation of the $v_\mathrm{sys}$ distribution: when assuming $\boldsymbol{v}_\mathrm{sys}$ is statistically isotropic (in other words, equally possible in all directions) for kinematically old pulsar systems, the $\boldsymbol{v}_\mathrm{sys}$ magnitude $v_\mathrm{sys}$ would follow a three-component Maxwell distribution with parameters identical to those of the three-component normal distribution (see Equation~\ref{eq:Vsys_distribution}). Furthermore, the analytical solution 
\begin{equation}
\label{eq:Vsys_CDF}
\mathcal{F}\left(v_\mathrm{sys}\right)=\sum_{i=1}^3 \alpha_i \left[\mathrm{erf}\left(\frac{v_\mathrm{sys}}{\sqrt{2}\gamma_i}\right)-\sqrt{\frac{2}{\pi}} \cdot \frac{v_\mathrm{sys}}{\gamma_i}\exp{\left(-\frac{{v_\mathrm{sys}}^2}{2\gamma_i^2}\right)}\right]
\end{equation}
can be derived for the cumulative distribution function (CDF) of the three-component Maxwell distribution, where erf is the error function, and the parameters $\alpha_i$ and $\gamma_i$ can be found in Table~\ref{tab:PDF_candidates}. 
The three-component Maxwell distribution as well as its corresponding CDF is displayed in Figure~\ref{fig:Vsys_PDF_CDF}.
In comparison, provided the same ``isotropic-$\boldsymbol{v}_\mathrm{sys}$'' assumption, no closed-form solution of $v_\mathrm{sys}$ can be obtained when each direction of $\boldsymbol{v}_\mathrm{sys}$ follows a Cauchy distribution. Therefore, we consider the three-component normal distribution a good and useful approximation to Equation~\ref{eq:vl_distribution} and the $v_\mathrm{l}$ PDF.

\section{Supplementary materials on dynamical population synthesis}
\label{ap:popSyn}

\subsection{Millisecond pulsar age distribution}
\label{subap:age_distribution}

\begin{figure*}[h]
    \centering    \includegraphics[width=0.65\textwidth]{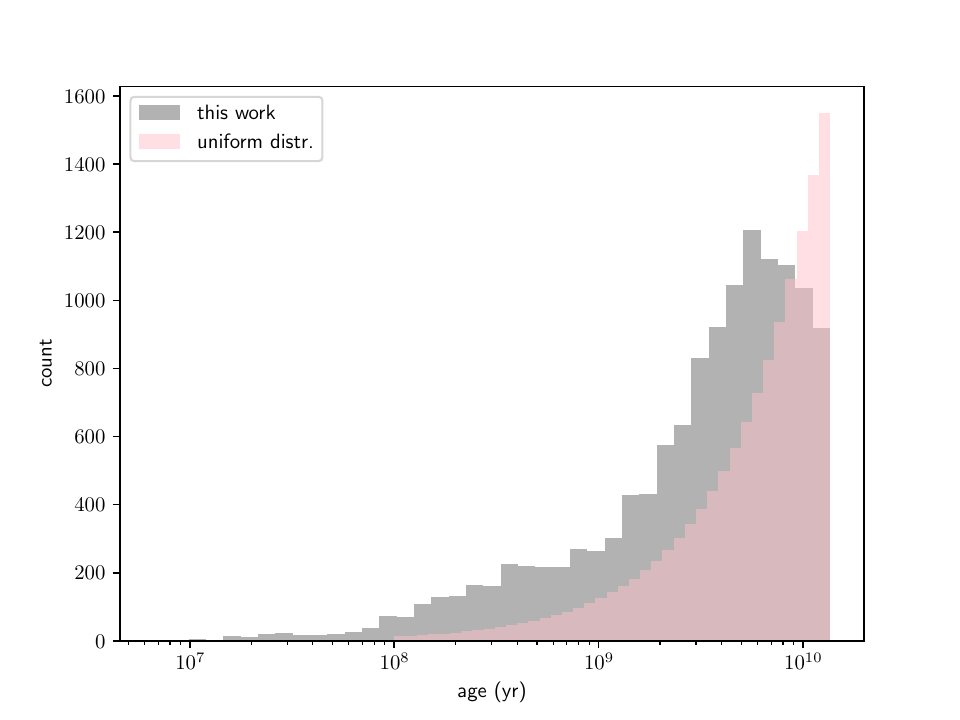}
    \caption{
    {\bf Gray:} The histogram of the millisecond pulsar age distribution adopted in DPS (see Appendix~\ref{subap:age_distribution}), with logarithmically spaced bins along the x-axis; {\bf Pink:} The uniform distribution $\mathcal{U}\left(0.1,13.6\right)$\,(Gyr) with logarithmically spaced bins along the x-axis, which is only used to examine the DPS robustness with respect to uncertainties in the MSP age distribution (see Appendix~\ref{subsubap:age_related_robustness}).}
    \label{fig:age_distribution}
\end{figure*}

The age distribution of MSPs is poorly constrained. We approached the rough MSP age distribution by the following procedure.
Firstly, we identified MSPs from the PSRCAT\textsuperscript{\ref{footnote:psrcat}} catalogue with the MSP criteria (i) and (ii) stated in Appendix~\ref{ap:sample_selection}. After removing sources without constraint on characteristic age $\tau_\mathrm{c}$, we obtained 399 MSP $\tau_\mathrm{c}$ values.
To derive the age distribution, we corrected the $\tau_\mathrm{c}$ sample by a sample of age correction factors (i.e., ratio between age and $\tau_\mathrm{c}$) estimated by \citet{Kiziltan10} for 42 MSPs having proper motion measurements (i.e., the reciprocals of ``$\tau_c/\tilde{\tau}'$'' reported in Table~1 of \citealp{Kiziltan10}). Specifically, we multiplied each of the $\tau_\mathrm{c}$ sample by each of the correction factor sample, which led to a sample of $399\times42=16,758$ MSP ages. After discarding ages larger than the age of the Galaxy (i.e., 13.6\,Gyr), we obtained a sample of 13,005 MSP ages (see Figure~\ref{fig:age_distribution} for the histogram of this sample).
We considered this age sample as an indicative approximation to the MSP age distribution, and adopted the age sample in the DPS analysis (see Section~\ref{subsec:popSyn_procedure}).

Despite the yet poorly constrained MSP age distribution, we find that the DPS results are insensitive to the uncertainties in the age distribution (see Appendix~\ref{subsubap:age_related_robustness}). 
While this insensitivity reinforces the robustness of the DPS results, it also prohibits the possibility of constraining age distribution of MSPs (and likely other kinematically old compact star systems) through astrometric studies.

\subsubsection{The robustness with respect to uncertainties in the age distribution}
\label{subsubap:age_related_robustness}

\begin{figure*}[h]
    \centering    \includegraphics[width=0.9\textwidth]{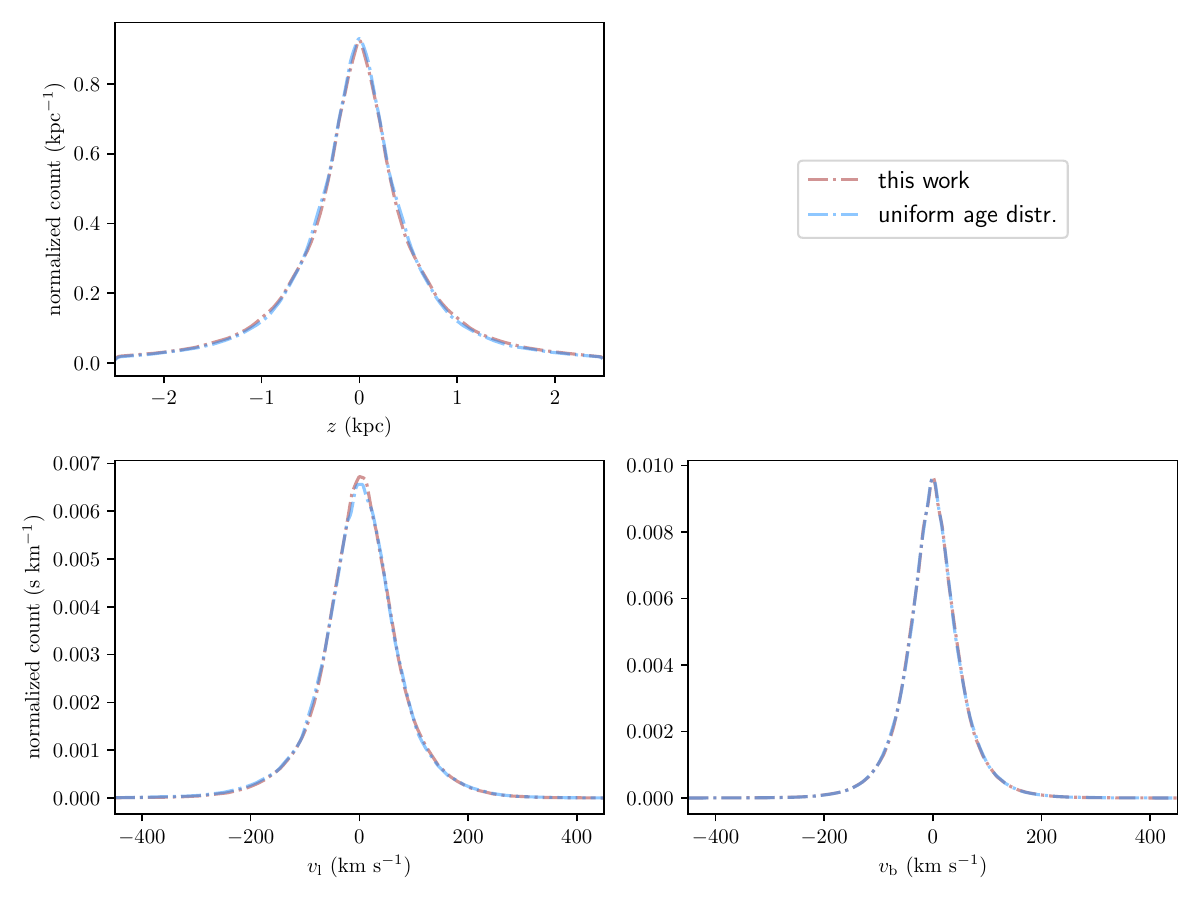}
    \caption{Difference between the smoothed histogram profiles of evolved $v_\mathrm{l}$, $v_\mathrm{b}$, and $z_\mathrm{e}$ 
    based on two distinct MSP age distributions --- the uniform distribution $\mathcal{U}\left(0.1,13.6\right)$\,(Gyr) in blue, and the MSP age distribution described in Appendix~\ref{subap:age_distribution}.}
    \label{fig:robustness_wrt_age_distribution_errors}
\end{figure*}

To investigate the impact of the uncertainties in the MSP age distribution on the DPS results, we reran DPS with the same setup except for the MSP age distribution. For the rerun, we adopted the uniform distribution $\mathcal{U}\left(0.1,13.6\right)$\,(Gyr) as the MSP age distribution (i.e., the ages are evenly distributed between 0.1\,Gyr and 13.6\,Gyr), which is overlaid to Figure~\ref{fig:age_distribution} in order to show its deviation from the age distribution of reference. Despite the difference between the uniform distribution and the reference distribution, we find that the two distinct age distributions render almost identical DPS results (see Figure~\ref{fig:robustness_wrt_age_distribution_errors}). Therefore, we conclude that the DPS results are robust with respect to the uncertainties in the MSP age distribution.

\subsection{Supplementary materials on the determination of the birth Galactic height distribution}
\label{subap:Zi_distribution}

\subsubsection{The underlying probability density function of the observed Galactic heights}
\label{subsubap:obs_z_PDF}

When searching for the candidates for the underlying PDF of the observed $z$ sample, we assumed that the PDF is both uni-modal and symmetric about zero. The same assumptions had been adopted in identifying the underlying PDF for the observed $v_\mathrm{l}$ (see Section~\ref{sec:v_sys_distribution}). We found that the one-parameter Cauchy distribution and the one-parameter Laplace distribution are both plausible candidates for the underlying PDF of the observed $z$ (see Figure~\ref{fig:v_and_z_distributions}). 
However, further analysis following the method detailed in Appendix~\ref{subap:Bayes_analysis} prefers the best-fit Laplace distribution over the Cauchy counterpart with $\ln{K}=7.8^{+0.7}_{-0.9}$, where $K$ is the Bayes factor.
The best-fit Laplace distribution for the observed $z$ sample has the scale height $\zeta^\mathrm{local}_\mathrm{obs}$ of 0.52\,kpc. We note that $\zeta^\mathrm{local}_\mathrm{obs}$ is derived from nearby field MSP system, which should be distinguished from the scale height $\zeta_1$ of Galaxy-wide field MSP systems.

\subsubsection{The determination of the Galaxy-wide birth scale height $\zeta_0$}
\label{subsubap:birth_scale_height}

\begin{figure*}[h]
    \centering    \includegraphics[width=0.85\textwidth]{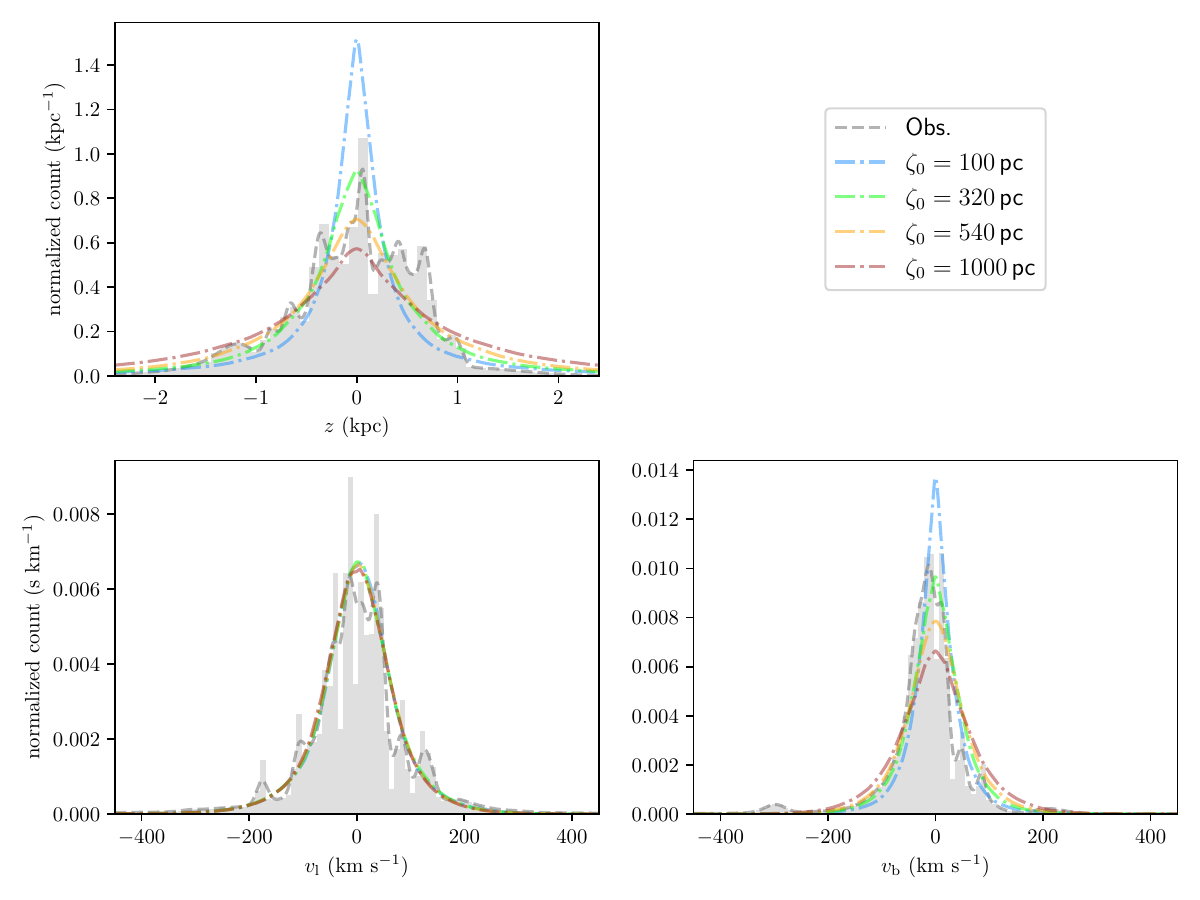}
    \caption{The smoothed histogram profiles of $z$, $v_\mathrm{l}$, and $v_\mathrm{b}$ derived from dynamical population synthesis based on different Galaxy-wide birth scale height $\zeta_0$ of field MSP systems. Overlaid are the visualized histograms (and their smoothed profiles) of the observed $z$, $v_\mathrm{l}$, and $v_\mathrm{b}$ (see Section~\ref{sec:obs}).}
    \label{fig:z0_dependence}
\end{figure*}

As mentioned in Section~\ref{subsec:initial_spatial_distribution}, various $\zeta_0$ would lead to different levels of the evolved $v_\mathrm{b}$ and $z_\mathrm{e}$.
Due to the complex distributions of the observed $z$ and $v_\mathrm{b}$ (see Figure~\ref{fig:v_and_z_distributions} or Figure~\ref{fig:z0_dependence}), we find it difficult to determine the $\zeta_0$ using statistical methods (e.g., Kolmogorov-Smirnov tests or the Wasserstein metric) that quantify the similarity between the observed samples and the DPS counterparts.
On the other hand, the evolved scale height $\zeta_\mathrm{e}^\mathrm{local}$ is expected to increase with $\zeta_0$ (see Figure~\ref{fig:z0_dependence}), and match the observed counterpart $\zeta^\mathrm{local}_\mathrm{obs}=0.52$\,kpc (see Appendix~\ref{subsubap:obs_z_PDF}). Therefore, we reran the DPS with incrementally larger $\zeta_0$ until the best-fit $\zeta_\mathrm{e}^\mathrm{local}$ reaches 0.52\,kpc.
In this way, we found $\zeta_0\approx0.32$\,kpc. The dependence of the evolved $v_\mathrm{b}$, $z_\mathrm{e}$ and $v_\mathrm{l}$ on $\zeta_0$ is illustrated in Figure~\ref{fig:z0_dependence}. As is shown in Figure~\ref{fig:z0_dependence},
both the evolved $v_\mathrm{b}$ and $z_\mathrm{e}$ match the observed counterparts at $\zeta_0=0.32$\,kpc.
Additionally, the average levels of both evolved $|\mathrm{b}|$ and $|z_\mathrm{e}|$ rise with $\zeta_0$ as expected.  
In comparison, the evolved $v_\mathrm{l}$ stays robust with respect to changes in the birth vertical distribution of field MSP systems.

Finally, we note that the observed $z$ sample might be biased by Galactic-latitude-related selection effects. As a major example, pulsars located at low Galactic latitudes $b$ are more susceptible to propagation effects caused by the ionized interstellar medium (IISM), which might reduce the detection rate of these low-$|b|$ pulsars.
This relative drop in the number of low-$|b|$ pulsars would become increasingly significant as further-away pulsars are searched and observed with large telescopes.  
In Section~\ref{subsec:binary_vs_solitary}, we found that the observed scale height of binary field MSPs is larger than solitary field MSPs, which may indicate the larger difficulty to detect binary MSPs (compared to solitary MSPs) in scattered low-$|b|$ sky regions.
On the other hand, most pulsar search programs have been focusing on low-$|b|$ sky regions, which serves as another selection effect that would increase the relative number of low-$|b|$ pulsars. 
We note that we did not take into account the impact of complex $b$-related selection effects on the observed $z$ (and $v_\mathrm{b}$) in this work.

\begin{figure*}[h]
    \centering    \includegraphics[width=0.7\textwidth]{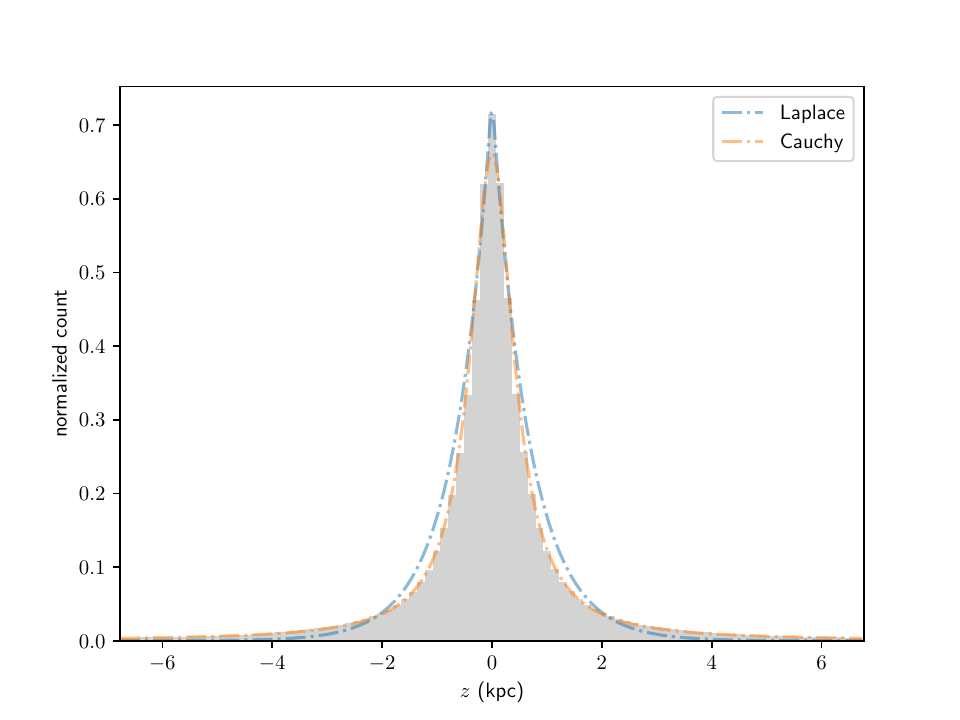}
    \caption{The histogram of Galaxy-wide evolved Galactic heights $z_\mathrm{e}$ derived from dynamical population synthesis. Overlaid are the best-fit Laplace and Cauchy distributions of the $z_\mathrm{e}$ sample, with the scale parameters of 0.68\,kpc and 0.48\,kpc, respectively.}
    \label{fig:Galaxy_wide_Ze_distribution}
\end{figure*}

\end{appendix}

\end{document}